\documentclass[a4paper,11pt]{article}

\usepackage{jheppub}
\usepackage[T1]{fontenc}
\usepackage{graphicx}
\usepackage{amsmath}
\usepackage{amsmath}
\usepackage{xspace}
\usepackage{subfig}
\usepackage{color}
\usepackage[utf8]{inputenc}
\usepackage{commath}
 \usepackage{cancel}
\usepackage{slashed}
\usepackage{fancyvrb}
\usepackage{multirow}
\newcommand{\calW}{{\cal W}}

\usepackage{xcolor}

\title{ Testing Neutrino Dipole Portal by Long-lived Particle Detectors at the LHC}

\author[a]{Wei Liu,}
\author[b]{Yu Zhang}

\affiliation[a]{Department of Applied Physics and MIIT Key Laboratory of Semiconductor Microstructure and Quantum Sensing, Nanjing University of Science and Technology, Nanjing 210094, China}
\affiliation[b]{School of Physics, Hefei University of Technology, Hefei 230601,China}

\emailAdd{dayu@hfut.edu.cn}

\abstract{
 We discuss the potential of using detectors aimed for searching long-lived particles~(LLP) at the high-luminosity LHC run, to probe the neutrino dipole models. This is achieved by taking the heavy neutral leptons~(HNL) of the models as candidates of the LLPs.
 Taking into account the  dipole couplings to the weak bosons, $d_{W,Z}$, which control the production of the HNLs at the LHC, we discuss the reach on the electromagnetic dipole couplings, $d_\gamma$, by searching for a single high-energy photon at LLP detectors. 
Four typical scenarios are considered in this paper, scenario A, B with $d_{W}=0$ or $d_{Z}=0$, and scenario C, D with $d_{W,Z}\gg d_\gamma$. We show the sensitivity on $d_\gamma$, can be fairly different depending on the relations between the $d_{W,Z}$ and $d_\gamma$.
And the LLP detectors can potentially extend the sensitivity on dipole couplings during the High-luminosity runs of the LHC in certain scenarios.
}
\begin{document}
\maketitle
\flushbottom
\setcounter{footnote}{0}

\section{Introduction}
\label{sec:intro}
The discovery of tiny neutrino masses, with non-explanation within the Standard Model~(SM) of the particle physics, is regarded as one of the most direct evidence points towards new physics beyond the SM. In efforts to explain the neutrino masses, additional right-handed neutrinos, also referred as the heavy neutral leptons~(HNL) $N$ are widely considered~\cite{Balaji:2019fxd,Balaji:2020oig,Delgado:2022fea,Barducci:2022gdv,Ding:2019tqq,Shen:2022ffi,Deppisch:2018eth,Deppisch:2019kvs,Liu:2022kid,Liu:2021akf,Liu:2022ugx,Beltran:2022ast,Zhou:2021ylt,Abada:2018sfh,Fernandez-Martinez:2022gsu,Abada:2022wvh,Arganda:2015ija,Bai:2022lbv,Das:2017nvm,Das:2012ze,Das:2015toa,Das:2016hof,Alok:2022pdn,Butterworth:2019iff}. They are singlets under the SM gauge groups. However, the HNLs can still interact with SM leptons $L$ and Higgs field~$H$ via a Yukawa interaction, $\mathcal{L} \supset NHL$, 
which accounts for the generation of the tiny Dirac neutrino masses. 

The experimental searches for such HNLs have received a lot attention, see Ref.~\cite{Abdullahi:2022jlv} for a recent review. Among them, the production of the HNL from the Yukawa interaction, so called the neutrino portal is widely considered.
New interactions to the $N$ can lead to novel signatures and features in their production and decay. For example, HNLs with gauge interactions are studied in Refs.~\cite{Deppisch:2018eth,Amrith:2018yfb,Deppisch:2019kvs,Liu:2021akf,Liu:2022kid,Liu:2022ugx}. In this work, we focus on another case, where the HNLs couple to the SM via the so-called diople portal, $\mathcal{L} \supset d \bar\nu_L \sigma_{\mu \nu} F^{\mu \nu} N$, where $F^{\mu \nu}$ stands for the electromagnetic field strength tensor, $d$ is the strength of magnetic dipole, and $\nu_L$ is the SM neutrino~\cite{Magill:2018jla}. This case is interesting, especially if the neutrino portal is subdominant.

The dipole portal models have been investigated at different existing experiments in various literature~\cite{Aparici:2009fh, Giunti:2014ixa, Aparici:2013xga, Coloma:2017ppo, Abazajian:2017tcc,Shoemaker:2018vii, Magill:2018jla, Brdar:2020quo, Plestid:2020vqf,Jodlowski:2020vhr, Schwetz:2020xra,  Ismail:2021dyp,Miranda:2021kre,Dasgupta:2021fpn,Atkinson:2021rnp,Kamp:2022bpt,Gustafson:2022rsz,Huang:2022pce,Li:2022bqr,Acero:2022wqg,Feng:2022inv,Hati:2022tfo,Mathur:2021trm,Bolton:2021pey,Ovchynnikov:2022rqj,Zhang:2022spf, Zhang:2023nxy,Ovchynnikov:2023wgg,Guo:2023bpo}. Ref.~\cite{Magill:2018jla} summarises the limits on the neutrino magnetic dipole based at colliders, beam-dump and neutrino experiments, astrophysics, cosmology, dark matter searches as well as future projection at the proposed SHiP detector. Ref.~\cite{Brdar:2020quo} revisits the limits at a neutrino or dark matter experiment by the detection of an upscattering event
mediated via a transition magnetic moment. Ref.~\cite{Plestid:2020vqf} discussed the possibility at experiments aiming for solar neutrinos. Ultrahigh energy neutrino telescopes can also be used to probe the dipole models, sensitive to very massive HNLs~\cite{Huang:2022pce}. 
Meanwhile, projections at other proposed future experiments are also investigated for  Forward LHC Detectors \cite{Jodlowski:2020vhr, Ismail:2021dyp}, Icecube  \cite{Coloma:2017ppo}, SuperCDMS \cite{Shoemaker:2018vii}, DUNE \cite{Schwetz:2020xra}, CE$\nu$NS and E$\nu$ES \cite{Miranda:2021kre}, as well as electron colliders \cite{Zhang:2022spf,Zhang:2023nxy,Ovchynnikov:2023wgg}. 

In most of scenarios considered by the existing literature, the dipole models can be simplified, only including the coupling $d_\gamma$between the sterile, active neutrinos and electromagnetic field strength tensor, as the energy scale is below the electroweak~(EW) scale. Nonetheless, if the energy possessed by the HNLs is comparable or even higher than the electroweak scale, e.g. HNLs produced at colliders, the SM gauge invariant dipole couplings $d_W$ and $d_Z$ should also be considered. 

In this work, we investigate the possibility where the HNLs are produced at the Large Hadron Collider~(LHC), and detected at the detectors aiming for searching long-lived particles~(LLP), including FASER~\cite{FASER:2018eoc}, MoEDAL-MAPP~\cite{Pinfold:2019nqj, Acharya:2022nik} and FACET~\cite{Cerci:2021nlb} at the high luminosity runs of the LHC~(HL-LHC). The beam-dump experiments can also be sensitive to the case where the HNLs are LLPs.
Comparing to the existing studies using beam-dump experiments, owing to the high energy scale at the LHC, the SM gauge invariant dipole couplings can play a crucial role. 
As we will shown in the rest of the paper, depending on the SM gauge invariant dipole couplings,  better sensitivity on the electromagnetic dipole couplings 
than the current limits can be yielded using LLP detectors.

We orangise the paper in the following order. In section~\ref{sec:model}, we briefly introduce the neutrino dipole portal model. The LLP detectors at the LHC is discussed at section~\ref{sec:detector}, followed by the investigation of their sensitivity for the dipole portal model at section~\ref{sec:sen}. And we conclude this paper in section~\ref{sec:con}.

\section{Neutrino Dipole Portal Model}
\label{sec:model}
The effective Lagrangian of the neutrino dipole $\mathcal{L} \supset d \bar\nu_L \sigma_{\mu \nu} F^{\mu \nu} N$ is only applicable at low energies.
The Lagrangian of the neutrino dipole, which respect the full gauge symmetries of the SM can be written as~\cite{Magill:2018jla}
\begin{equation}
	\mathcal{L} \supset \bar{L}(d_{\calW}^k \calW_{\mu \nu}^a \tau^a + d_B^k B^{\mu \nu}) \tilde{H}\sigma_{\mu \nu} N+\mathrm{H.c.},
\end{equation}
$\tilde{H}=i\sigma_2H^*$ and $\tau^a=\sigma^a/2$, where $\sigma^a$ is the Pauli matrix.
In this form, it can describe the new physics beyond the EW scale.

After spontaneous symmetry breaking, the Lagrangian becomes
\begin{equation}
\mathcal{L} \supset d_W^k(\bar{\ell^k} W^-_{\mu \nu} \sigma_{\mu \nu} N) + \bar{\nu}_L^k(d_\gamma^k F_{\mu \nu}-d_Z^k Z_{\mu \nu}) \sigma_{\mu \nu} N +\mathrm{H.c.}.
\label{eq:LWB}
\end{equation}
Hence, the right-handed neutrinos $N$ couple to SM photon, $Z$ and $W$ bosons via the dipole couplings $d_\gamma^k$,$d_Z^k$, and $d_W^k$ respectively. 

For a given lepton flavor $k$, the dipole couplings $d_\gamma^k$,$d_Z^k$, and $d_W^k$ in the broken phase are linearly dependent by only two parameters $d_\calW$ and $d_B$ in the unbroken phase, such that~\footnote{The superscript $k$ of the lepton flavor is omitted in the rest of the paper to simplify the notation, otherwise stated.}
\begin{eqnarray}
d_\gamma&=&\frac{v}{\sqrt{2}}\left(d_B\cos\theta_{w}+\frac{d_{\calW}}{2}\sin\theta_w\right), \nonumber\\ 
d_Z&=&\frac{v}{\sqrt{2}}\left(\frac{d_{\calW}}{2}\cos\theta_{w}-d_B\sin\theta_w\right), \nonumber\\ 
d_W&=&\frac{v}{\sqrt{2}}\frac{d_{\calW}}{2}\sqrt{2}.
\end{eqnarray}
By further assuming $d_{\calW}=a\times d_B$, we have
\begin{eqnarray}
d_Z &=&\frac{d_\gamma(a\cos\theta_w-2\sin\theta_w)}{2\cos\theta_w+a\sin\theta_w}, \nonumber\\
d_W &=& \frac{\sqrt{2}a d_\gamma}{2\cos\theta_w+a\sin\theta_w} .
\end{eqnarray}

{The above expressions are only true if the effective field theory~(EFT) was valid at the LHC.
The dipole couplings $d_{\calW, B}$ are dim-6 operators, while
$d_{\gamma, Z, W}$ are generated after spontaneous symmetry breaking, so are dim-5 operators. The EFT should be valid with the largest $d_\gamma \sim \frac{v}{\Lambda^{2}} \sim \frac{100~ \text{GeV}}{\Lambda^{2}}$~\cite{Racco:2015dxa, Magill:2018jla}, and $d_\gamma \sim \frac{1~ \text{GeV}}{\Lambda^{2}}$ in the perturbative limit.} 
In our following calculation,  since the production of $N$ mainly comes from  on-shell decay of the $W/Z$ at the LHC, the EFT is valid as long as the cutoff scale $\Lambda \gtrsim M_{W/Z}$ which indicates that  the $d_\gamma$ can be as large as $\mathcal{O} (10^{-(3-4)})$.

{The dipole couplings can be connected to the generation of the neutrino masses via loop diagrams, if a Majorana mass term $\mathfrak{m}_N$ exists. However, in this paper, we consider the HNL as purely Dirac fermion, or quasi-Dirac  with a small Majorana-type mass splitting satisfying $\mathfrak{m}_N\ll m_N$ \cite{Magill:2018jla}. Large dipole couplings can still be compatible to the observed tiny neutrino masses, since they are decoupled, therefore as free parameters.}

Thus, we have three independent free parameters in our model
\begin{eqnarray}
(m_N, d_\gamma, a),
\end{eqnarray}
where $m_N$ is the mass of the HNL.

\section{Signals of the HNLs at the LHC}
\label{sec:detector}
\subsection{Production and Decay of the HNL}
\begin{figure}[t!]
\centering
\includegraphics[width=0.49\textwidth]{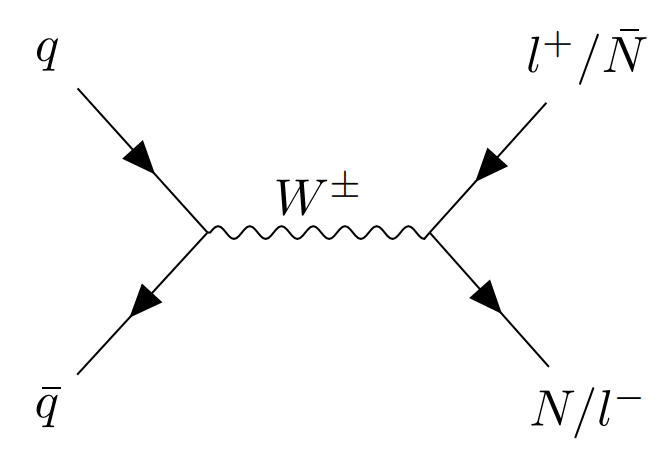}
\includegraphics[width=0.49\textwidth]{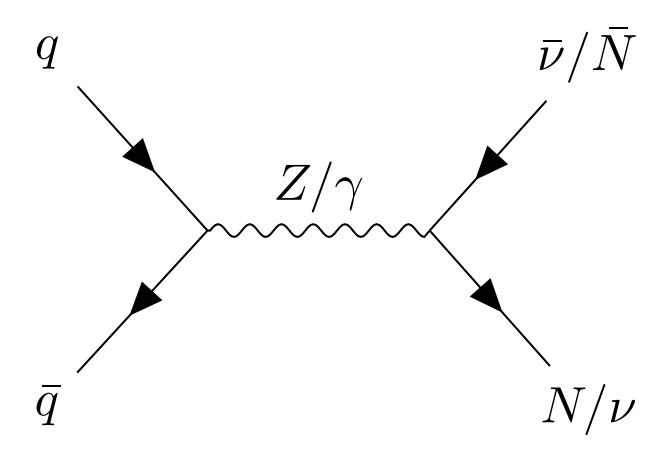}
\caption{The Feynman diagrams of the production of the right-handed neutrinos $N$ at the LHC.}
\label{fig:feyn}
\end{figure}
We consider the HNL at the LHC produced by the decay of the gauge bosons, i.e. $pp \rightarrow W^{\pm} \rightarrow N l^{\pm}$, and
$pp \rightarrow Z, \gamma \rightarrow N \nu$, as shown in Fig.~\ref{fig:feyn}~\footnote{Although $N$ are Dirac particles, we omit the sign of them as well as their decay products, unless stated.} \footnote{{The HNL can also be produced via meson decays, see Ref.~\cite{Jodlowski:2020vhr, Barducci:2022gdv}. However since we focus on the interplay between the different dipole couplings, and the meson decays channels are dominantly controlled by the $d_\gamma$,
therefore we only consider production of HNL via gauge boson decays, and leave the meson decay channels in future work.}}. {The production of the HNL via gauge boson decays can also be triggered by the active-sterile neutrino mixings. Nevertheless, if the neutrino masses were generated via type-I seesaw, the active-sterile neutrino mixings should be tiny, thus this contribution can be negligible, }
The production cross section depends on the couplings of $N$ to the gauge bosons, $d_W$, $d_Z$ and $d_\gamma$ as well as $m_N$, therefore by $(m_N, d_\gamma, a)$. The $N$ subsequently decays via the same couplings, with the decay width
\begin{eqnarray}
\Gamma_{N \rightarrow \nu \gamma} &=& \frac{|d_\gamma|^2 m_N^3}{4 \pi} ,\\
\Gamma_{N \rightarrow \nu Z} &=& \frac{|d_Z|^2 (m_N^2-M_Z^2)^2 (2m_N^2+M_Z^2)}{8 \pi m_N^3} \Theta(m_N > M_Z),\\
\Gamma_{N \rightarrow  W \ell } &=& \frac{|d_W|^2}{8 \pi m_N^3} \sqrt{(m_N^2-(M_W-m_\ell)^2(m_N^2+(M_W-m_\ell)^2))} \\\nonumber
&\times& (2 m_\ell^2(2m_\ell^2-4m_N^2-M_W^2)+(m_N^2-M_W^2)(2m_N^2+M_W^2))\Theta(m_N > M_W+m_\ell).
\label{eq:decay}
\end{eqnarray}
 $N$ can also decay via off-shell $W$ and $Z$~\cite{Atre:2009rg, Bondarenko:2018ptm},
\begin{eqnarray}
\Gamma_{N \rightarrow  \text{2body} } &\propto& |d_{W,Z}|^2 \frac{G_F^2 m_N^3 f_M^2}{10 \pi},\\
\Gamma_{N \rightarrow  \text{3body} } &\propto& |d_{W,Z}|^2 \frac{G_F^2 m_N^5}{100 \pi^3} ,
\end{eqnarray}
where $G_F$ and $f_M$ are Fermi constant and meson decay width, respectively.
As we focus on the $N$ which can lead to LLP signals at the LHC, for most of the parameter space with  $m_N \lesssim 2$~GeV, we only have appreciable $\Gamma_{N \rightarrow \nu \gamma}$, hence ${\rm Br}(N \rightarrow \nu \gamma)\simeq\text{100 \%}$ and $\Gamma(N) \propto |d_\gamma|^2$.

Having understood the expressions of the production and decay of the $N$, Monte-Carlo simulations are performed to analyse the kinematics. We use the Universal FeynRules Output~(UFO)~\cite{Alloul:2013bka, Degrande:2011ua} of the neutrino dipole model developed in Ref.~\cite{Zhang:2022spf}, which is fed to the event generator {\tt MadGraph5aMC$@$NLO} -v2.6.7~\cite{Alwall:2014hca} to generate events at parton level. Shower, hadronization, etc are handled by {\tt PYTHIA v8.306}~\cite{Sjostrand:2014zea}. Detector level simulation and the clustering of the events by later purpose is performed by {\tt Delphes v3.5.0}~\cite{deFavereau:2013fsa} and {\tt FastJet v3.2.1}~\cite{Cacciari:2011ma}, respectively.

\begin{figure}[t!]
\centering
\includegraphics[width=0.49\textwidth]{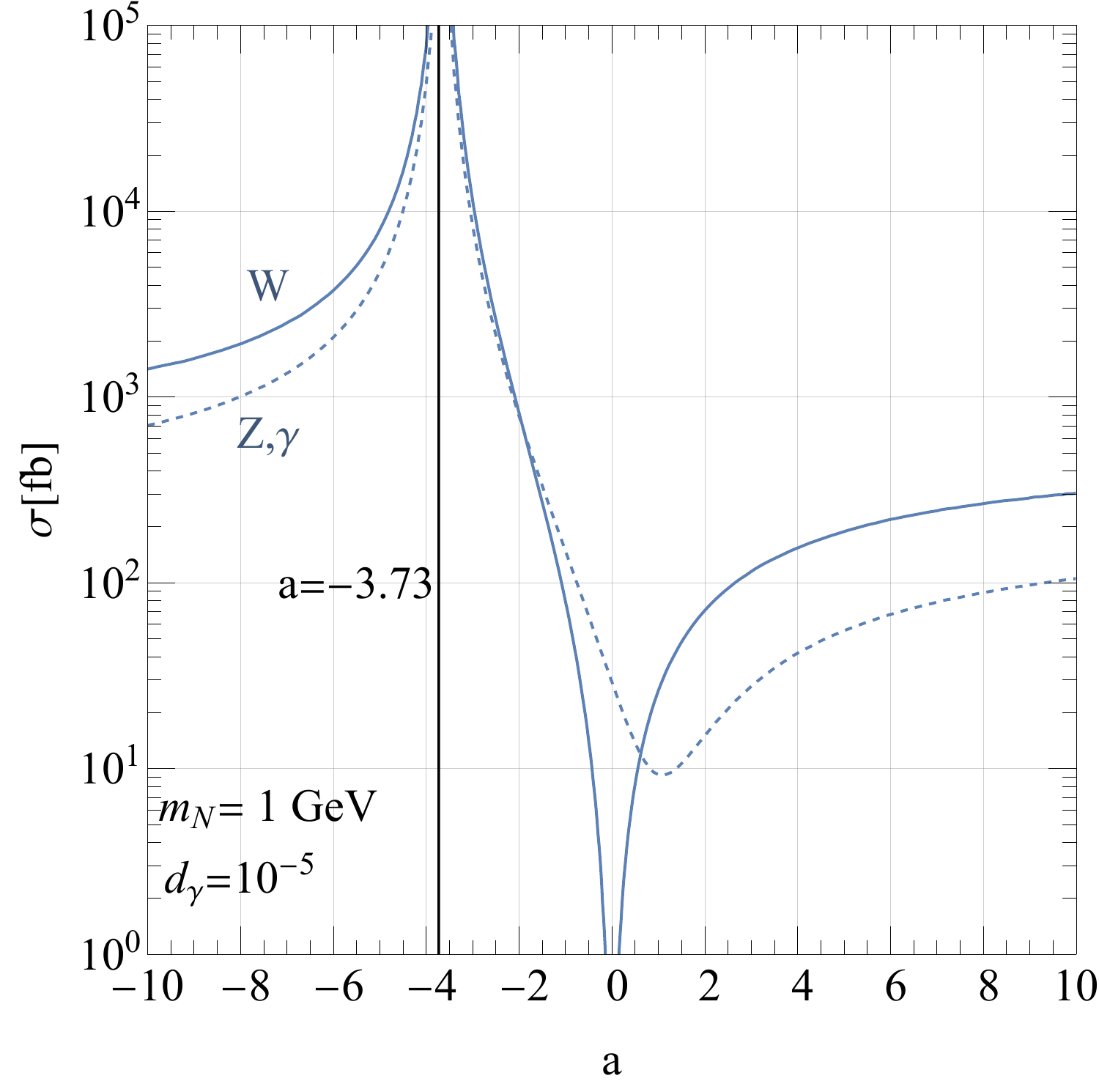}
\includegraphics[width=0.49\textwidth]{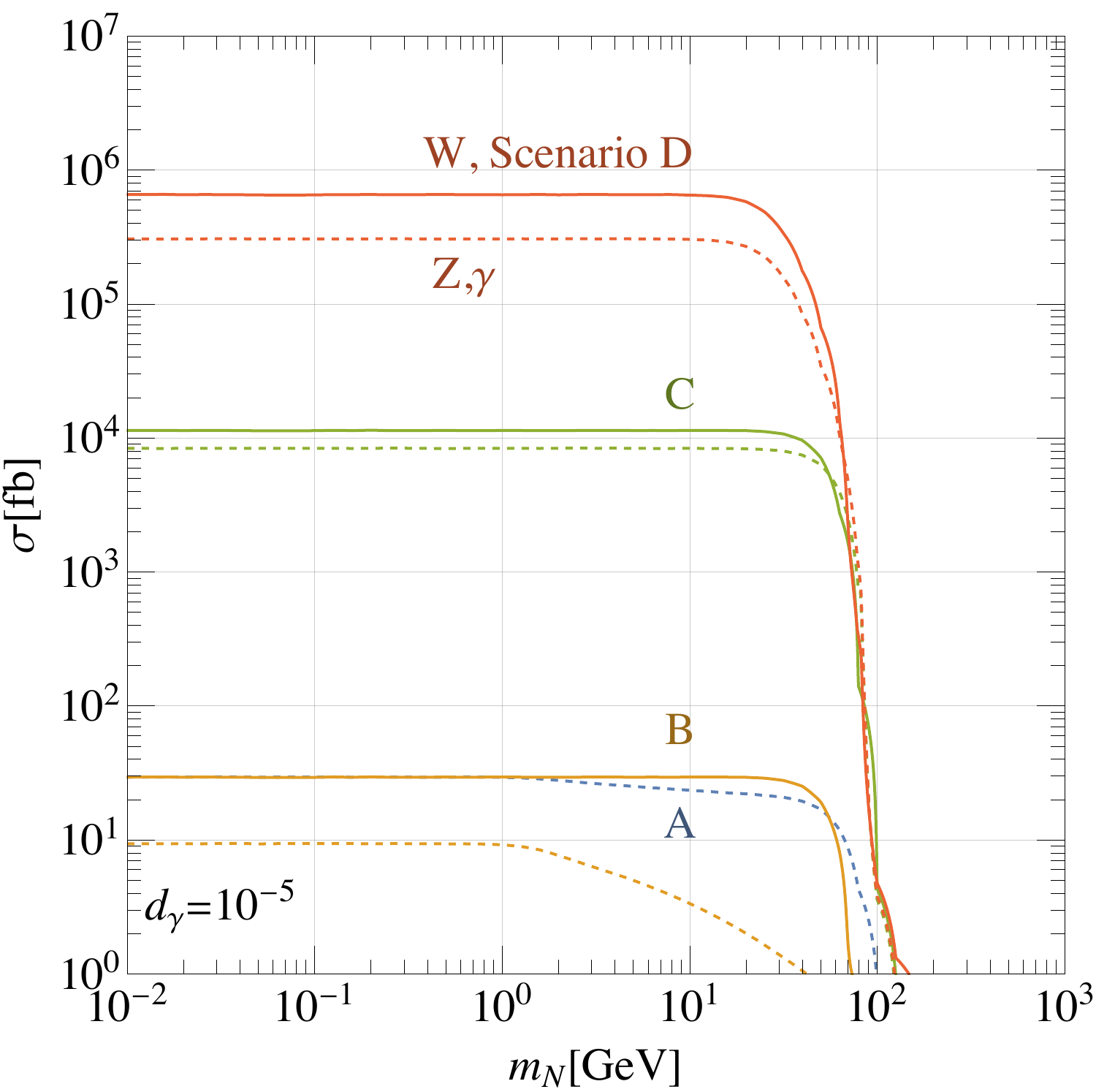}
\caption{Left: The cross section of the processes $pp \rightarrow W^{\pm}  \rightarrow N l^{\pm}$ ({solid}), and  $pp \rightarrow Z,\gamma  \rightarrow N \nu $ ({dashed}) at the 13 TeV LHC as a function of $a$, when $d_\gamma = 10^{-5}$ and $m_N = 0.1$ GeV. Right: Same but as a function of $m_N$ when $d_\gamma = 10^{-5}$ for Scenario A~($a=0$), B~($a=2 \tan \theta_w$), C~($a=-3$), and D~($a=-3.73$).}
\label{fig:funa}
\end{figure}

The cross sections of the processes $pp \rightarrow W^{\pm}  \rightarrow N l^{\pm} $~(blue line), and  $pp \rightarrow Z/\gamma  \rightarrow N \nu$~(orange line) at the 13 TeV LHC as a function of $a$ when $d_\gamma = 10^{-5}$ and $m_N = 0.1$ GeV, are shown in Fig.~\ref{fig:funa} left. It is clear that the cross sections depend strongly on $a$. For the $W$ mediated processes, they are only controlled by $d_W$, which has a singularity with $a = -2 \cot \theta_w \approx -3.73$. 
 {Whereas their cross section becomes vanishing when $a$ approaches zero leading to $d_W\sim 0$.} The $Z,\gamma$ mediated processes have shown similar behavior, only they get minimum cross section where $a = 2 \tan \theta_w$ with $d_Z =$0. The minimum is non-vanishing since the $\gamma$ mediated processes still exist.

To this end, we select four typical scenarios to reflect the dependence on the high energy couplings $d_W$ and $d_Z$, where $a=0$ for Scenario A, and $a=2 \tan \theta_w, -3$ and -3.73 for Scenario B, C and D, respectively, as summarised in Table.\ref{tab:scen}.  We further show the dependence on the HNL mass $m_N$ for the two scenarios in Fig.~\ref{fig:funa} right {with $d_\gamma = 10^{-5}$}. For Scenario A, $W$ mediated processes vanish, while $Z,\gamma$ mediated processes can still get a constant value about $30$ fb when $m_N < M_W$, and drop off gradually to below 1 fb when $m_N$ approaches  {100 GeV}. Things becomes different when look at Scenario B, now the $Z$ mediated processes vanishes, the $N \nu$ final states can still be produced via $\gamma$ with only $\sim 10$ fb cross section. The $W$ mediated processes have similar cross section comparing to the $Z$ ones for Scenario A. As for the Scenario C and D,
now $W$ mediated processes get larger cross section than $Z/\gamma$, reaches $\mathcal{O}(10^{4,5} )$ fb, while dropping sharply to below 1 fb when $m_N$ approaches 100 GeV. And the $Z,\gamma$ mediated processes have similar behavior.

\begin{table}[]
	\centering
	\begin{tabular}{c|c|c} \hline\hline
		Scenario &  Assumptions   &  Relations\\ \hline\hline
		A &  $d_{\calW}=0$   & $d_Z=-d_\gamma \tan\theta_w$; $d_W=0$\\ \hline
    B &  $d_{\calW}=2\tan\theta_w\times d_B$ & $d_Z=0$, $d_W=\sqrt{2}d_\gamma\sin\theta_w$ \\ \hline 
  C &  $d_{\calW}=-3 \times d_B$ & $d_Z\approx11.250 \times d_\gamma$, $d_W \approx 13.258 \times d_\gamma $  \\ \hline 
		D &  $d_{\calW}=-3.73 \times d_B$ &  $d_Z \approx 139.55 \times d_\gamma $;  $d_W \approx 173.52 \times d_\gamma $ \\\hline\hline
	\end{tabular}
	\caption{The four scenarios we taken in this paper.}
	\label{tab:scen}
\end{table}

\begin{figure}[t!]
\centering
\includegraphics[width=0.49\textwidth]{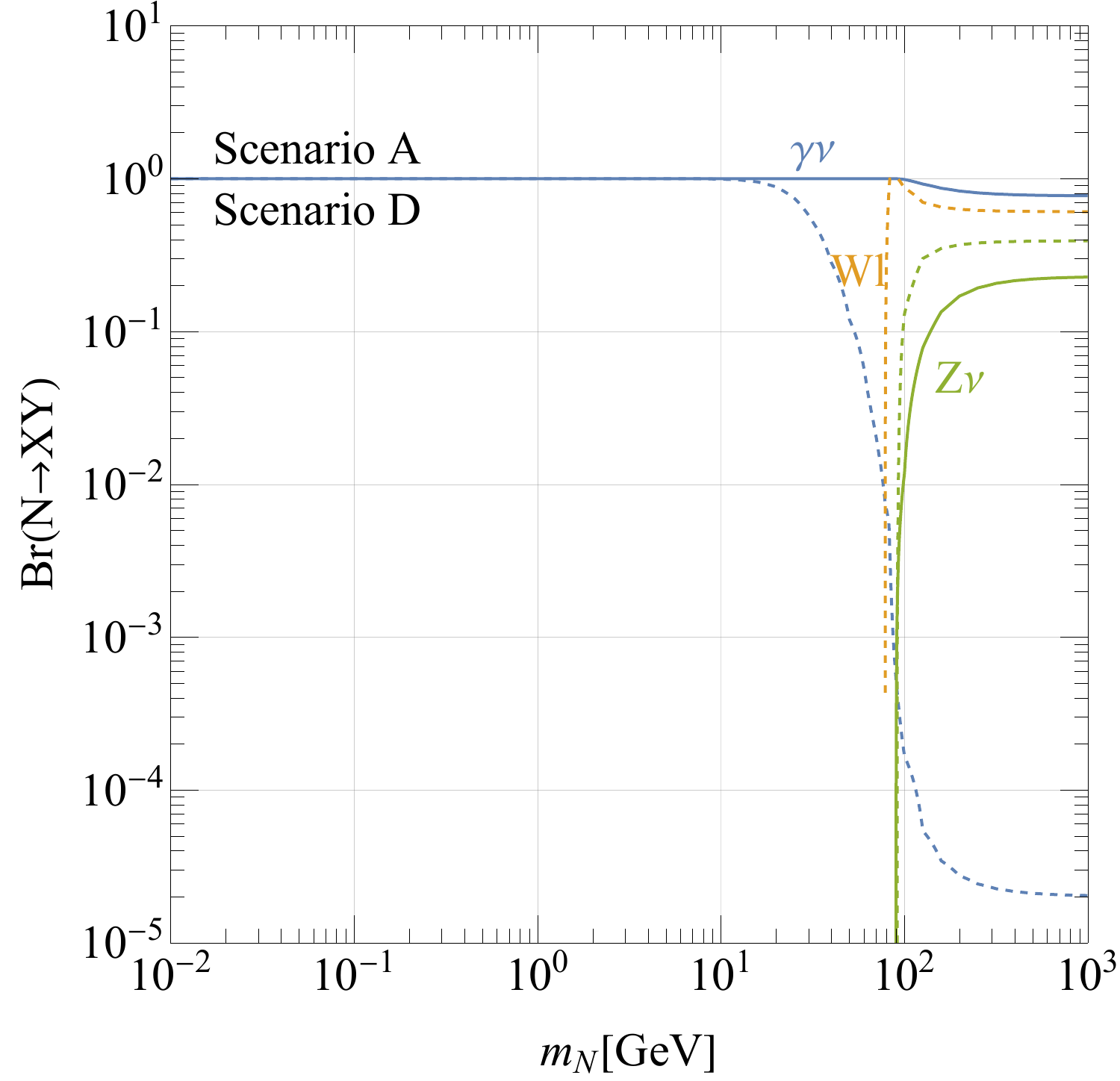}
\includegraphics[width=0.49\textwidth]{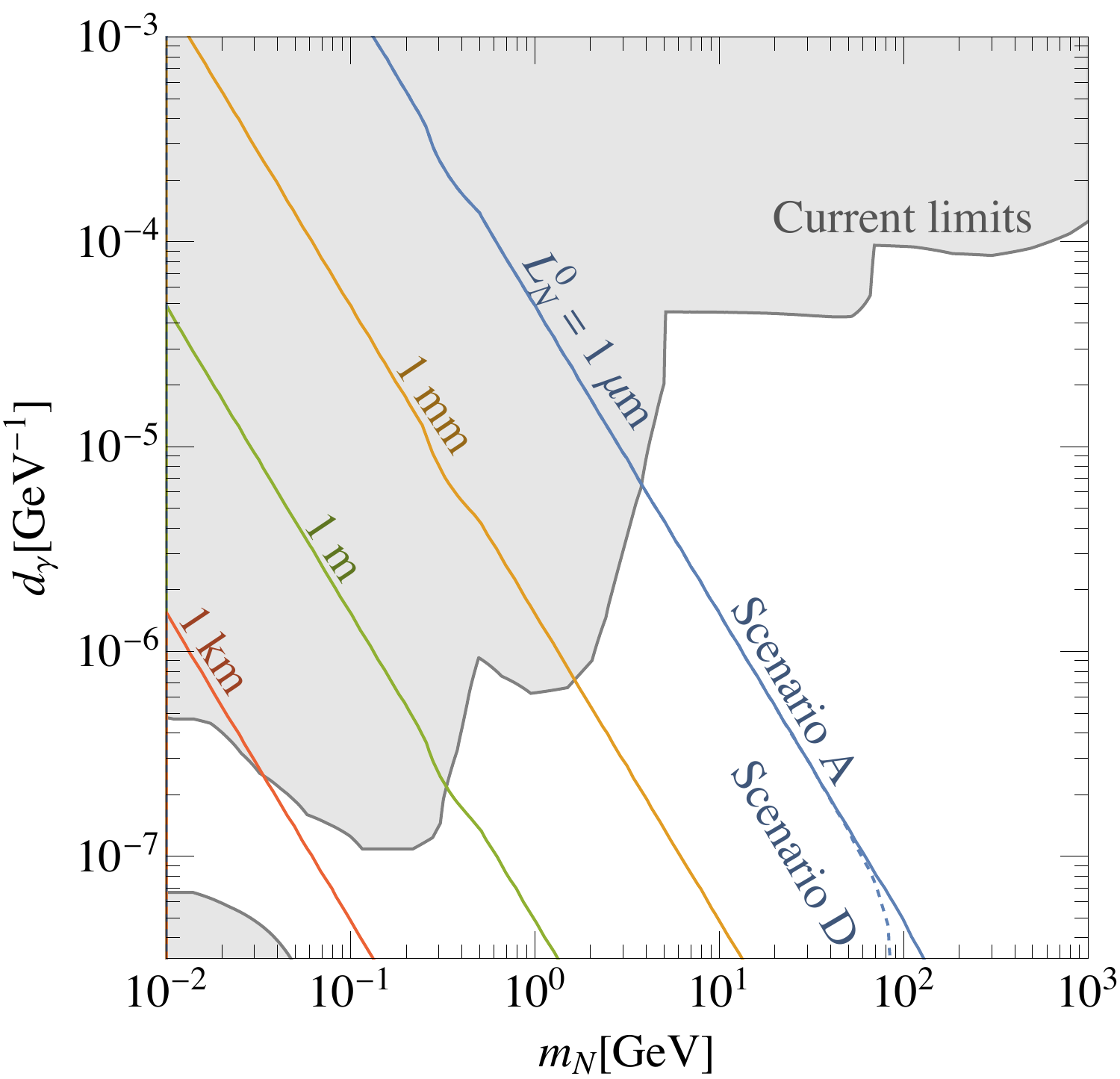}
\caption{Left: ${\rm Br}(N \rightarrow X Y)$ as a function of $m_N$ for Scenarios A and D. Right: Proper decay length of the HNL in ($m_N$,$d_\gamma$) plane. The solid~(dashed) lines correspond to Scenario A~(D). Current limits from Ref.~\cite{Magill:2018jla,Jodlowski:2020vhr} are overlaid for comparison, only for Scenario A.}
\label{fig:fdecay}
\end{figure}

In Fig.~\ref{fig:fdecay}, we present the radiative decay branching ratio ${\rm Br}(N \rightarrow \nu \gamma)$ as a function of $m_N$ for Scenarios A and D. We only show these two scenarios, since 
Scenario B and C are similar to A and D, respectively.
It can be found that in Scenario A there always be ${\rm Br}(N \rightarrow \nu \gamma)\simeq 1$ until $m_N > M_Z$ in which the decay channel into on-shell $Z$ boson $N\to Z\nu$ opens. Whereas in Scenario D,  the radiative decay branching ratio starts to decrease  rapidly from $m_N\gtrsim 10$ GeV, since the decays via an off-shell $W,Z$ become sizeable. 
Due to the large {ratio of $d_{W,Z}/d_\gamma$ } for Scenario D, ${\rm Br}(N \rightarrow \nu \gamma)$ is vanishing once $m_N > M_{W}$, opposite to Scenario A where it is still appreciable. And decays into on-shell $W,Z$ become the dominant channels.
We show the proper decay length of HNL, $L_N^0$ in ($m_N$,$d_\gamma$) plane. Current limits from Ref.~\cite{Magill:2018jla, Jodlowski:2020vhr} are overlaid for Scenario A, while the limits for Scenario D will be shown later. From the figure, we obtain a useful analytical approximation of $L_N^0$ for $m_N \ll M_W$ no matter what value of $a$,
\begin{eqnarray}
L_N^0 \approx 2.5~\text{cm} \times \left(\frac{d_\gamma}{10^{-5}}\right)^{-2} \times \left( \frac{m_N}{0.1~\text{GeV}}\right)^{-3}.
\label{eq:length}
\end{eqnarray}
It is evident to find that under current limits, the HNLs can have decay length of $\mathcal{O}$(m), which means they can be regarded as candidates of LLPs. 
The difference between the two scenarios in decay length do not enter into the parameter space interesting for LLPs consideration where $m_N < 10$ GeV, as shown that the decay length are only different between Scenario A and B when $L_N^0 \lesssim 10^{-6}$~m.

This is important for the following analyses of the LLP signals. To generate macroscopic decay length of one particle for it to become a LLP, feeble interactions are required. If the LLPs are produced and decayed via the same interactions, this will leads to insignificant signal events in most cases. Nevertheless, in the model we consider, the production is controlled by $d_{Z,W,\gamma}$ or $(a, d_{\gamma})$, whereas the decay does not depend on $a$ in our consideration of LLP signals.
This means that without making the $N$ not long-lived anymore, the production rates of $N$ at LHC can be larger depending on the value of $a$ in our model.

\subsection{Analyses for the Long-lived Particle Detectors at the LHC}
Bear that in mind, we proceed the detailed analyses for LLP signals in this section. 
Although there exists quite a lot searches for LLPs at the multi-purpose detectors at the LHC, i.e. ATLAS, CMS and LHCb, no signatures of LLPs are found so far~\cite{Alimena:2019zri}. 

Benefited from their large distance to the interaction point~(IP) and shields to stop the SM final states, specialized detectors aimed at probing LLPs
might lead to more positive prospect of the discovery of the LLPs.
 Among them, the FASER and MoEDAL-MAPP detectors are already installed and operated since Run-3 of the LHC. The FASER detector is located about 480 meters away from the IP of the ATLAS experiment, at a very forward direction. The MoEDAL-MAPP~(MAPP) detector is a new subdetector of the MoEDAL experiment, which is located about 50-100 meters away from the IP of the LHCb.  In the meantime, other designs of LLP detectors such as AL3X~\cite{Gligorov:2018vkc}, ANUBIS~\cite{Bauer:2019vqk}, CODEX-b~\cite{Gligorov:2017nwh}, FACET~\cite{Cerci:2021nlb} and MATHUSLA~\cite{Curtin:2018mvb} detectors are also proposed. A short review for all of these detectors can be found in Ref.~\cite{Abdullahi:2022jlv}. Considering the proposed detectors, we focus on the ones which can reconstruct the photon signals, including FASER, MAPP and FACET\cite{Barducci:2022gdv}.
 We take FACET to compare with FASER, since they are both at the forward direction. We focus on the phase two designs of the FASER~(FASER-2) and MAPP~(MAPP-2) detectors at the HL-LHC, since they have larger geometrical coverage and luminosity, providing optimistic reach of the LLP signals. FACET are also considered to be operated at the HL-LHC. We summarise the geometrical coverage and luminosity for the detectors we considered in Table~\ref{tab:LLP}~\footnote{The original MAPP detector actually has a ring-like shape, here we roughly consider it as a cuboid to
simplify the calculation.}.

\begin{table}
	\centering
	\begin{tabular}{|l|c|c|c|c|c|c|}
		\hline
		Detectors & $L_x$~[m]  & $L_y$~[m] & $L_{xy}$~[m] & $L_z$~[m] & Luminosity~[fb$^{-1}$]\\
		\hline
		FASER-2  & $-$  & $-$  & $[0,1]$ & $[475, 480]$ &  3000 \\
		\hline
		MAPP-2  & $[3,6]$  & $[-2,1]$  & $-$ & $[48, 61]$ & 300  \\
		\hline
  	FACET  & -  & -  & [0.18, 0.5] & $[101, 119]$ & 3000  \\
		\hline
	\end{tabular}
	\caption{The geometrical coverage and luminosity corresponding for FASER-2~\cite{FASER:2018eoc}, MAPP-2~\cite{Pinfold:2019nqj, Acharya:2022nik}, and FACET~\cite{Cerci:2021nlb}.   }
	\label{tab:LLP}
\end{table}

The expected number of the observed events at these LLP detectors can be expressed as
\begin{eqnarray}
N_{\text{signal}}/\mathcal{L} \approx \sigma(pp \rightarrow W/Z, \gamma \rightarrow N \ell/\nu)  \times \epsilon_{\text{kin}} \times \epsilon_{\text{geo}} ,
\label{eq:nsignal}
\end{eqnarray}
here $\mathcal{L}$ is the integrated luminosity.  $\epsilon_{\text{kin, geo}}$ are the efficiencies due to the trigger requirement, and geometrical acceptance, respectively. 
{A kinematic threshold, $E_{vis} > 100$ GeV is put for FASER-2, following Ref.~\cite{Jodlowski:2020vhr}.}

{At FASER-2, for such high energies, the background can be suppressed. The main background for this single high-energy photon can be induced by the neutrino and muon. The neutrino-induced background can be cut away by the use of a dedicated preshower detector. While the muon-induced background can be vetoed by detecting the accompanying time-coincident muon~\cite{Jodlowski:2020vhr,FASER:2018bac}. It still remains to be difficult to estimate the number of residual background events in a reliable way, and it has beyond the scope of our current study. Therefore, we only show the results with fixed number of signal events for each detectors. $N_{\text{signal}}=3,~30$ is going to be shown for FASER-2, as the background has been discussed in detailed. The information of the background at FACET and MAPP-2 is not yet provided yet in literature. }

The geometrical acceptance is estimated as follows. In principle, $\epsilon_{\text{geo}}$ is related to the probability of the HNL to decay inside the detector volume, which is
a function of the momentum $p$, angle to the beam line $\theta$, and lab frame decay length $L_N^\text{lab}$, such as~\cite{FASER:2018eoc} 
\begin{eqnarray}
\mathcal{P}(p,\theta) = \left( e^{-(L-\Delta)/L_N^\text{lab} }- e^{-L/L_N^\text{lab}} \right) \Theta(R-\tan\theta L) \approx \frac{\Delta}{d} e^{-L/L_N^\text{lab}}  \Theta(R- \theta L) \ ,
\label{eq:decayinvolume}
\end{eqnarray}
where $\Theta$ is the Heaviside step function, $L$, $R$, and $\Delta$ are the distance to the IP, radius in the $xoy$ plane and length of the detector. $L_N^\text{lab}=c\tau\beta\gamma = c \tau p/m$ is the lab frame decay length of the LLP, where $c \tau$ is the proper decay length. 
However, Eq.~\ref{eq:decayinvolume} requires $L$ and $R$, being constants for different $\theta$, so it is only applicable for detectors like FASER-2 and FACET placed at a very forward direction and symmetric around the beam line. For MAPP-2, which have more complicated shape, we apply Monte-Carlo methods by inverse sampling of the cumulative distribution function according to the lifetime of the HNL.

\begin{figure}
\centering
\includegraphics[width=0.49\textwidth]{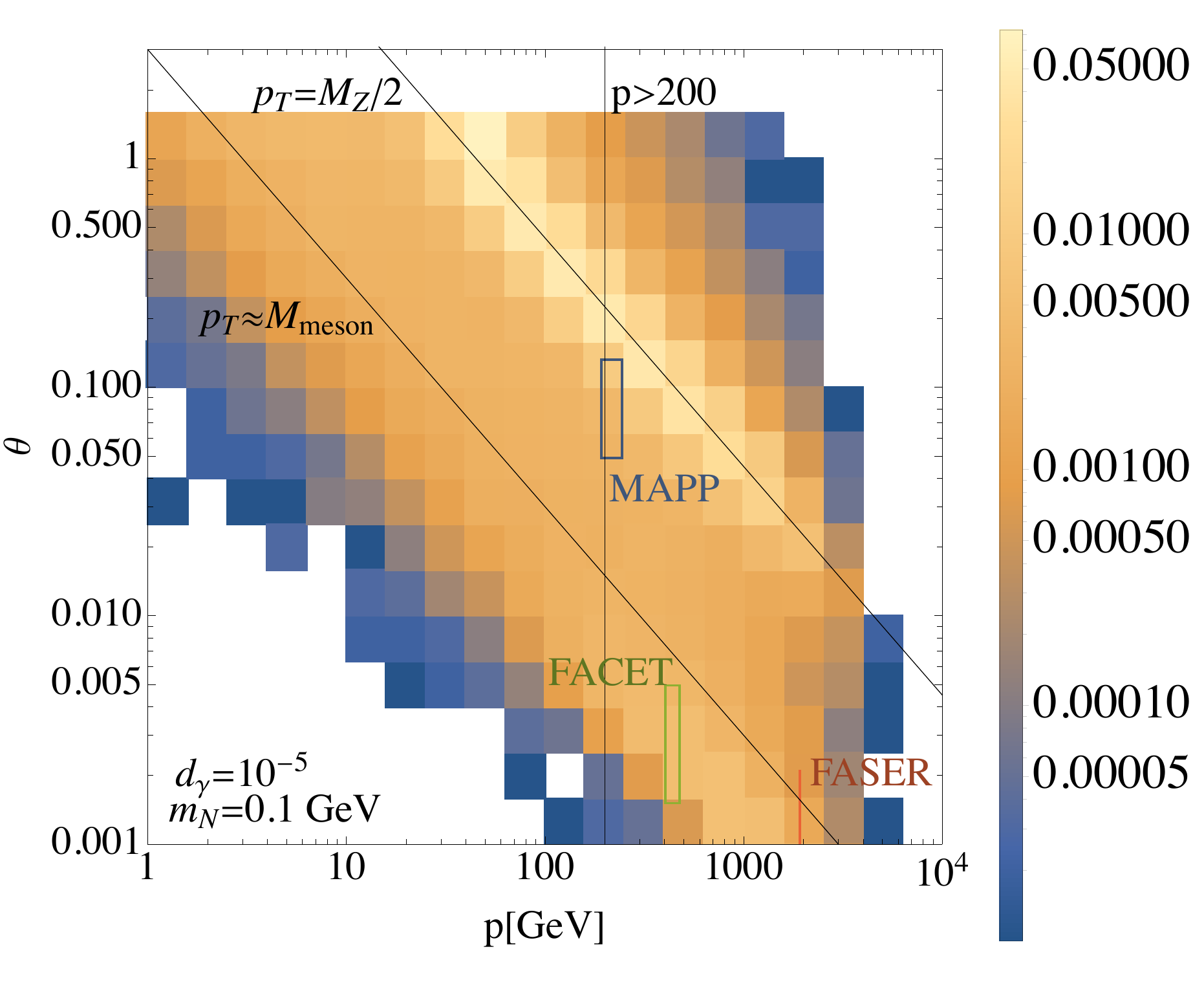}
\includegraphics[width=0.49\textwidth]{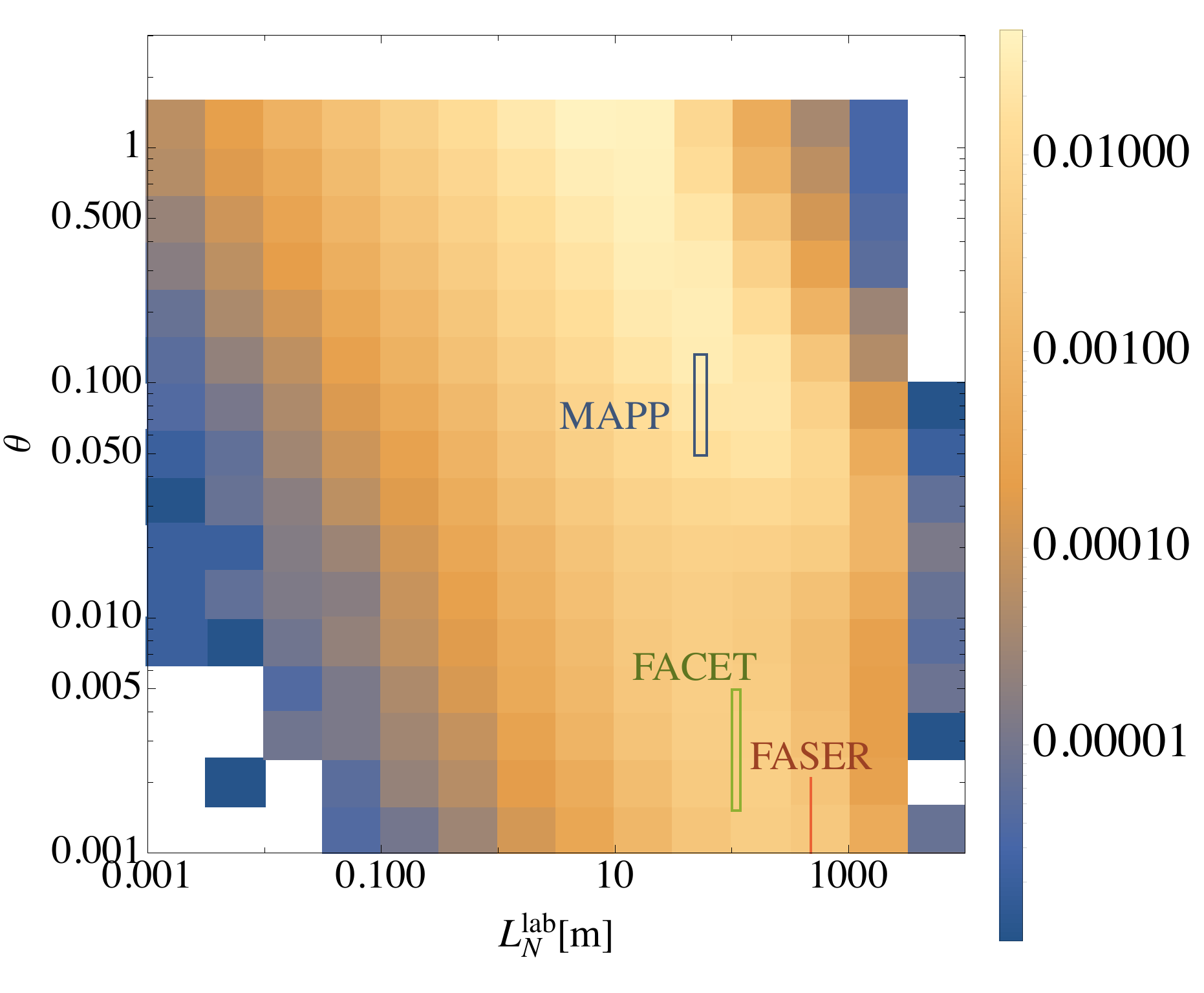}
\caption{In Scenario A, the distribution of the $p$ and $\theta$~(left) for $10^5$ events, as well as $L_N^\text{lab}$ and $\theta$~(right) for $10^6$ events of the HNLs from $pp \rightarrow W/Z,\gamma \rightarrow N \ell/\nu$ process. The approximate coverage of the FASER-2~(red), MAPP-2~(blue), and FACET~(green) detectors is overlaid for comparison. The colours represent the weight of each bin, which is normalised to one.
We fix $m_N =$ 0.1 GeV, and $d_\gamma = 10^{-5}$.}
\label{fig:pdis}
\end{figure}

\begin{figure}
\centering
\includegraphics[width=0.49\textwidth]{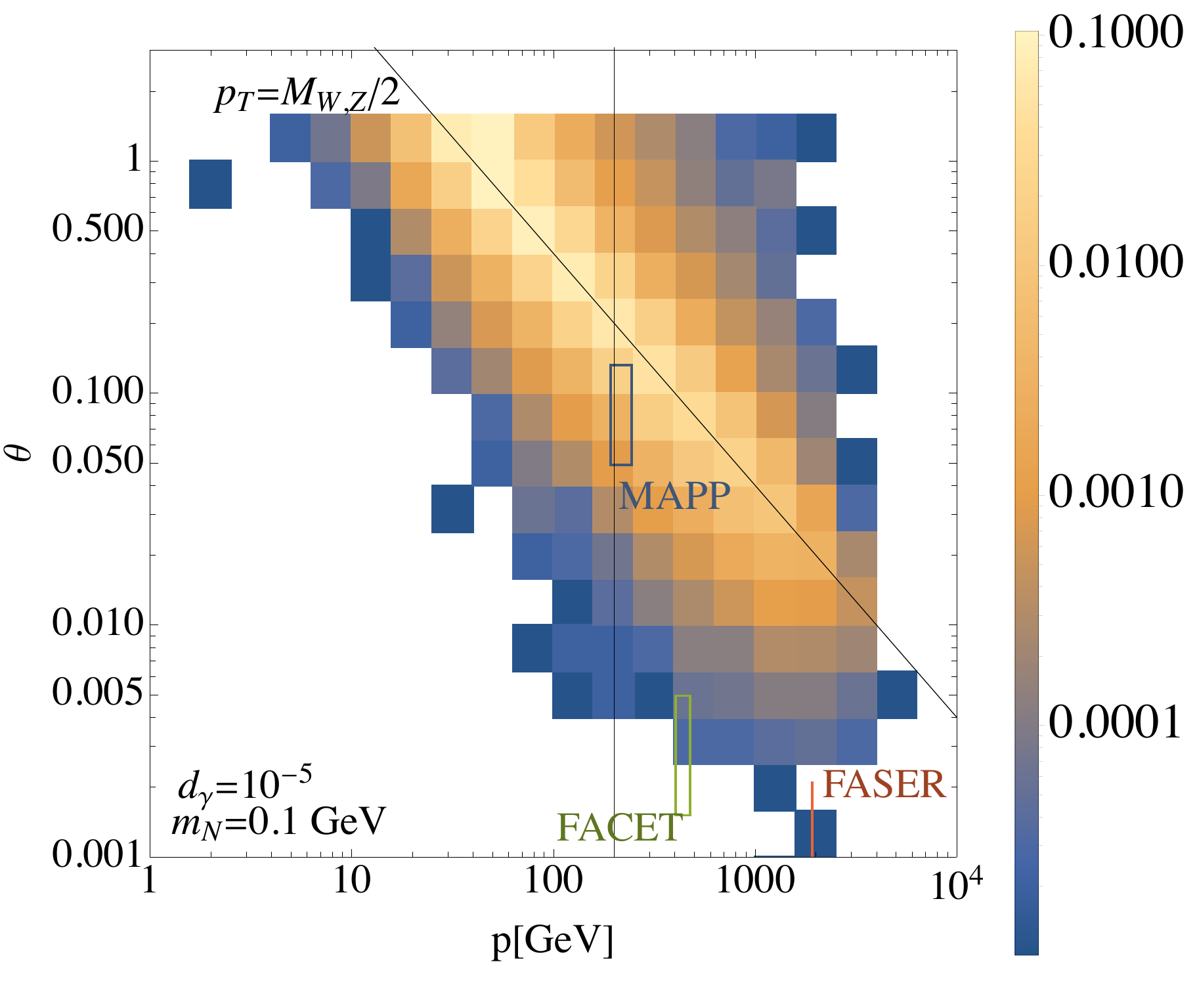}
\includegraphics[width=0.49\textwidth]{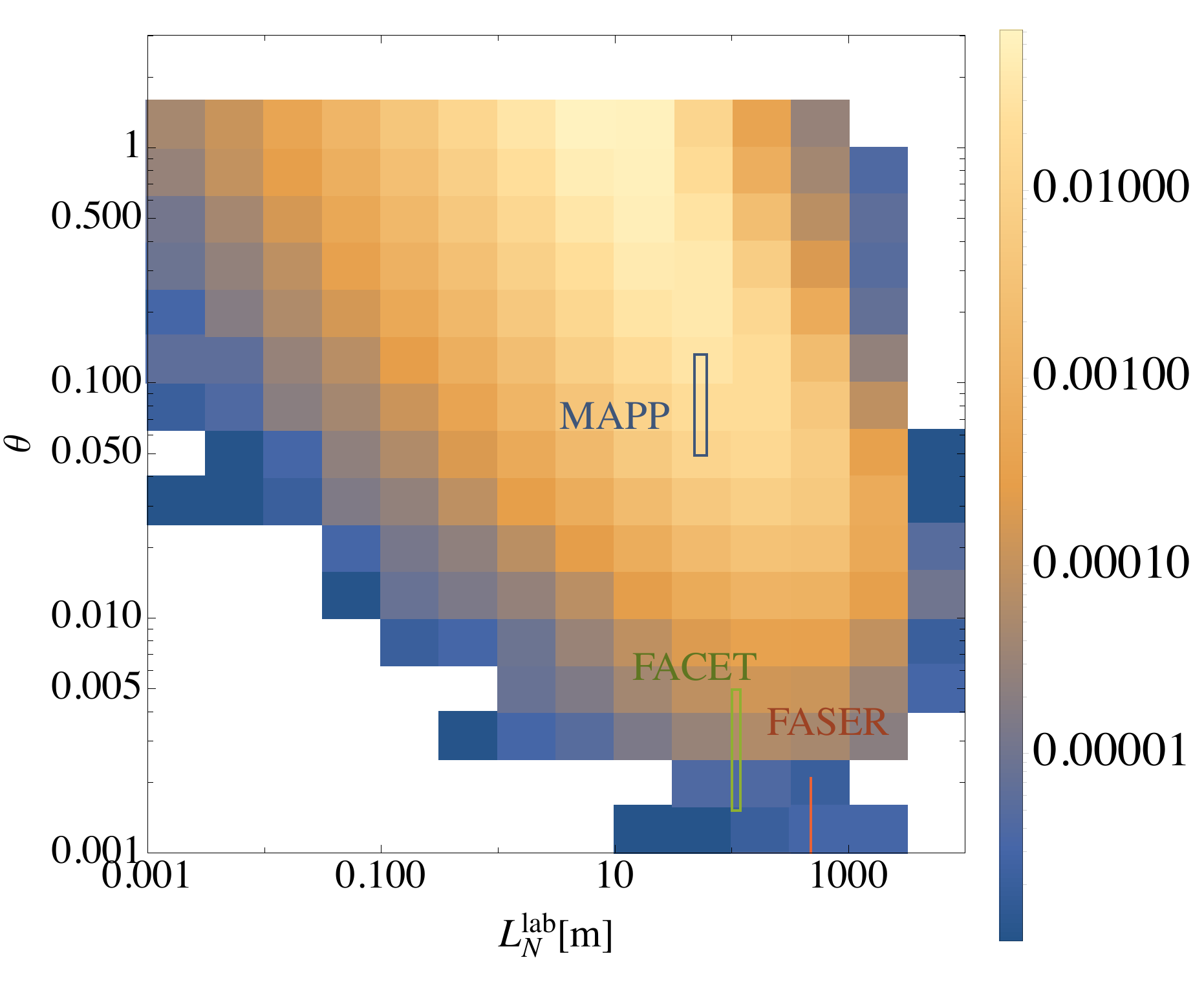}
\caption{The same but for Scenario D. }
\label{fig:pdisB}
\end{figure}

To roughly illustrate how the probability varies for different detectors, we show 
the distribution of the momentum $p$, angle to the beam line $\theta$, and lab frame decay length $L_N^\text{lab}$ for the HNLs in Fig.~\ref{fig:pdis} and~\ref{fig:pdisB} , at one benchmark where $m_N =$ 0.1 GeV and $d_\gamma = 10^{-5}$ for Scenario A and D, respectively. Again, Scenario B and C are similar to A and D, therefore not shown.
The approximate coverage of the FASER-2~(red), MAPP-2~(blue), and FACET~(green) detectors is overlaid for comparison. Nonetheless, the coverage on the $\phi$~($xoy$ plane) is not been shown, thus the resulting geometrical acceptance should be smaller comparing to the ones estimated from the figure.

In Fig.~\ref{fig:pdis}~left, we show the distribution of $p$ and $\theta$ of the HNLs for $10^5$ events in Scenario A. As shown in Eq.~\ref{eq:length}, the proper decay length $L_N^0$ is about $ 2.5~ \text{cm}$ for this benchmark. The lab frame decay length equals to $L_N^0 \times p/m_N$, therefore each detector requires the $p$ to be inside certain range to make the HNLs likely to decay within its volume. Nevertheless, the HNLs can still decay inside the detector volume for other values of $L_N^\text{lab}$, since their decay follow exponential distribution, but the probability is rather low. 
Both the $Z$ and $\gamma$ mediated processes contribute to the distribution for Scenario A. The distribution from $Z$ mediated process peaks around the line where $p_T = M_{Z}/2$
, since the transverse momentum of the $N$ is approximately half the mass of the mother particle $Z$ for a 1$\rightarrow$2 process, when
$m_N \ll M_{Z}$. However, for $\gamma$ mediated process, the distribution peaks where $p_T = p_T(\gamma)/2$, which can come from the remnant of the mesons masses, therefore covers a broader parameter space, especially for low $\theta$ region.
Among these detectors, MAPP-2 located the closet to the peak of the $Z$ mediated distribution. 
Whereas FACET and FASER-2 are located too far away from the $Z$ peak, but benefited from the coverage of low $\theta$ of the $\gamma$ mediated distribution, therefore can still obtain appreciable acceptance.

The effects of the trigger can be seen in Fig.~\ref{fig:pdis}~left, , i.e. $p > 200$ GeV from 
~$E_{vis}>$ 100 GeV, as $E_{vis}\approx p/2$ since both photon and neutrino are almost massless. At this benchmark, we can see that this trigger does not result in any difference, since the requirement for the HNLs to decay inside detector volume already ask them to be energetic enough. Especially, $p \sim 2$~TeV is needed for FASER-2.
However, when discuss other parameters, the proper decay length can be larger, so lower Lorentz factor subsequently lower $p$ of the HNLs are required. Since $L_N^0 \propto d_\gamma^{-2} \times m_N^{-3}$, so the momentum required $p \propto d_\gamma^{2} \times m_N^{3}$. For instance, when $m_N = $~0.1 GeV, if $d_\gamma = 10^{-6}$ instead of $10^{-5}$, FASER-2 now requires $p \sim 20$~GeV, which makes the $p>$ 200 GeV trigger effective to cut almost all the events. Generally speaking, trigger effects for the kinematical efficiencies $\epsilon_{\text{eff}}$ make the lowest $d_\gamma$ the detectors can reach larger, i.e. worse sensitivity. For a $p>$ $p_\text{low}$ trigger, the lowest $d_\gamma$ becomes $\sqrt{p_\text{low}}$ times larger, and about one magnitude for the $p>$ 200 GeV trigger.

In Fig.~\ref{fig:pdis}~right, we show the distribution  of $L_N^\text{lab}$ and $\theta$ of the HNL for Scenario A. This figure is quite similar to the left one, only the $x$ axis is scaled with a factor of $0.25$~m $\times$ GeV$^{-1}$, and the $L_N^\text{lab}$ contains exponential distribution since each $N$ decays exponentially. For each HNL, we simulate 10 events for the exponential distribution, so the statistics is higher, reaching $10^6$ events. Due to the exponential distribution, the distribution is modified, the parameter space far away from the peak now gets the tail from the exponential distribution. For example, FASER-2 now locates inside the bins with weight about $10^{-2}$, which is larger from Fig.~\ref{fig:pdis}~left.
It severs as a more direct view of the geometrical acceptance of these detectors.

Comparing the distribution between Scenario A and D with Fig.~\ref{fig:pdis} and~\ref{fig:pdisB}, both the distribution of the momentum $p$, angle to the beam line $\theta$, and lab frame decay length $L_N^\text{lab}$ has shown appreciable difference. The contribution from $\gamma$ mediated process is insignificant in Scenario B since its cross section are much lower than the ones mediated by $W$ and $Z$, therefore the distribution only surround where $p_T = M_{W,Z}/2$. 
For Scenario B, as shown in Fig.~\ref{fig:pdisB}, now FASER-2 and FACET locate too far away from the peak, only get the tail of the exponential distribution. On the other hand, MAPP-2 are closer to the peak, thus still covers similar weight of events as in Scenario A.

\begin{figure}
\centering
\includegraphics[width=0.49\textwidth]{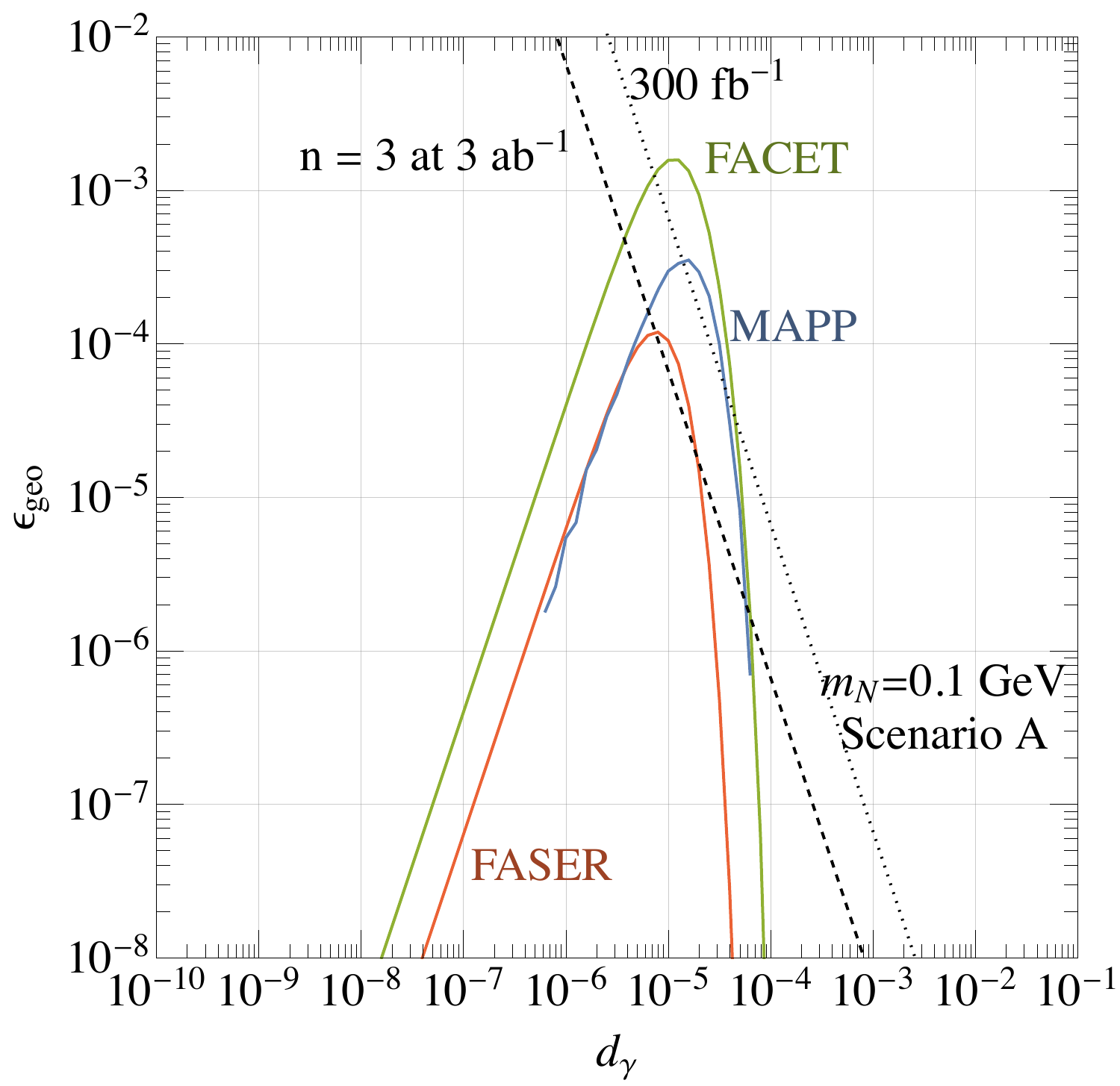}
\includegraphics[width=0.49\textwidth]{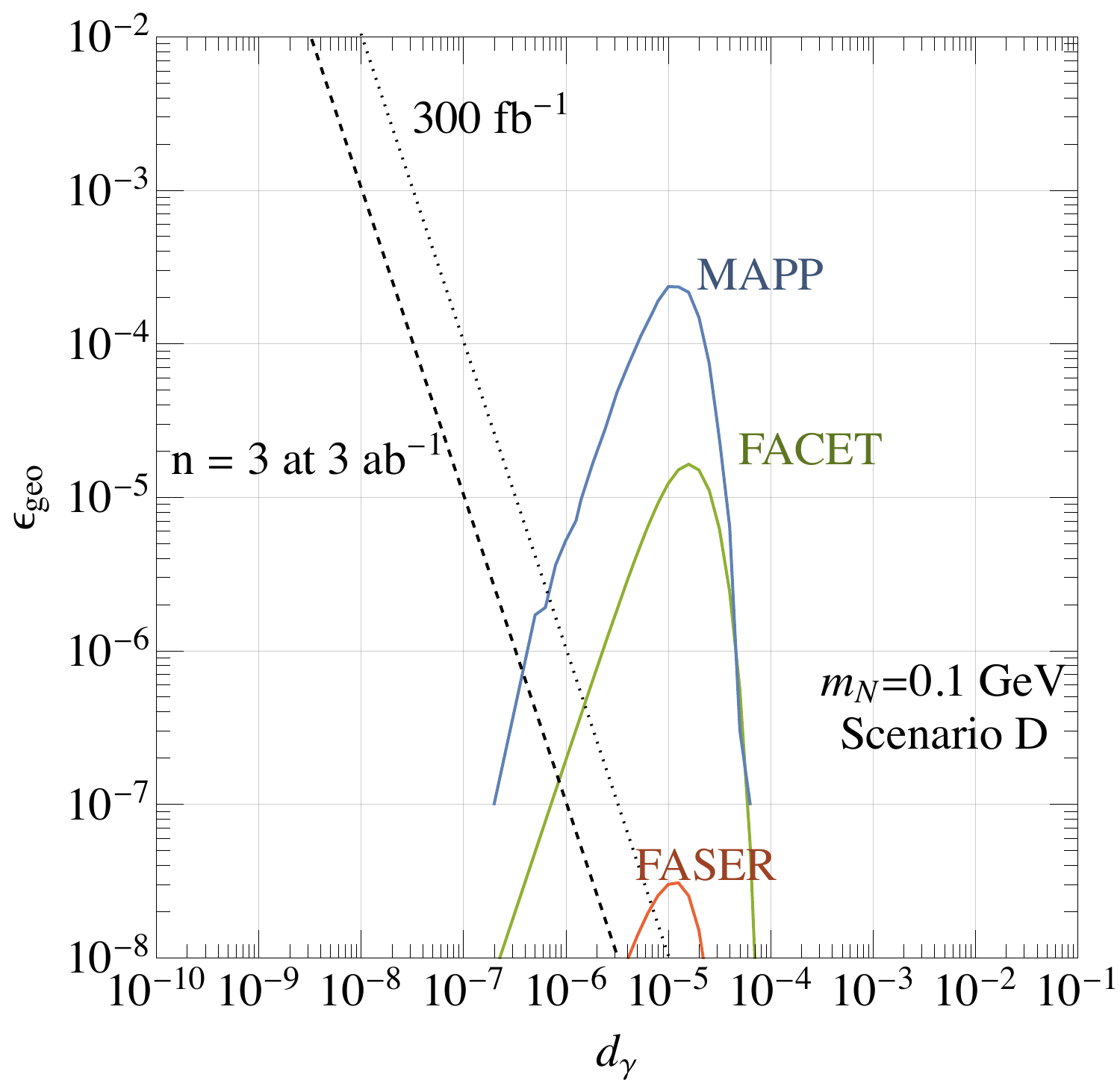}
\caption{The geometrical efficiencies of the aforementioned detectors for Scenario A~(left) and D~(right). The $\epsilon_{\text{geo}}$ required to make $N_{\text{signal}}=$~3 for Scenario A and D is demonstrated as the dashed~(dotted) black lines at 3000~(300) fb$^{-1}$ luminosity.
We fix $m_N =$ 0.1 GeV.}
\label{fig:feff}
\end{figure}

 We refer to Fig.~\ref{fig:feff} for the detailed geometrical acceptance $\epsilon_{\text{geo}}$ of each detector at the same benchmark for Scenarios A and D. 
When $d_\gamma = 10^{-5}$ and $m_N =$ 0.1~GeV, the geometrical acceptance $\epsilon_{\text{geo}}$ is about $10^{-4}$ for MAPP-2, $10^{-3~(-5)}$ for FACET, and $10^{-4~(-7)}$ for FASER-2, in Scenario A~(D). 
 This is smaller as than it shown up in Fig.~\ref{fig:pdis} and~\ref{fig:pdisB}~right, and FASER-2, MAPP-2 as well as FACET only gets small fraction of bins in Fig.~\ref{fig:pdis}~right.
 The difference between Scenario A and D, is from the different contribution of the $\gamma$ mediated process. The $\gamma$ mediated process can lead to appreciable distribution of HNLs for low $\theta$ as shown in Fig.~\ref{fig:pdis}~left, therefore FASER-2 and FACET get larger acceptance in Scenario A where the contribution of this process is significant.
 The number of signal events 
 $N_{\text{signal}}$ can be obtained from Eq.~\ref{eq:nsignal}. $\sigma(pp \rightarrow W/ Z, \gamma \rightarrow N \ell/\nu) $ is about $(d_{\gamma}/10^{-5})^2 \times 10^{1(6)}$~fb, when $a = 0~(-3.73)$ for Scenario A~(D) 
and $m_N = 0.1$~GeV from Fig.~\ref{fig:funa}~left.

At the HL-LHC, with 3000~(300) fb$^{-1}$ integrated luminosity for the IP of FACET and FASER-2~(MAPP-2), the $\epsilon_{\text{geo}}$ required to make $N_{\text{signal}}=$~3 for Scenario A and D are demonstrated as the dashed black lines. Below the lines, the detectors suffer in low geometrical acceptance, leading to low signal events and vice versa. The range of $d_\gamma$ to make $N_{\text{signal}} > $~3 can be estimated from the intersection points of the $\epsilon_{\text{geo}}$ curves of the detectors and the $N_{\text{signal}}=$~3 lines.
For Scenario A, when $m_N =$ 0.1~GeV, we get $d_\gamma \gtrsim  10^{-5~(-6)}$ for 
FASER-2 and MAPP-2~(FACET) detectors. For Scenario D,  
we have $d_\gamma \gtrsim 10^{-6}$ in order to make $N_{\text{signal}}>$~3 for 
FASER-2, MAPP-2 and FACET.

\section{Results}
\label{sec:sen}
Now we show the sensitivity at the HL-LHC. According to the Lagrangian in Eq.~\ref{eq:LWB}, $d_\gamma$ can vary for different lepton flavours $k$, where $k = e, \mu, \tau$. Several existing limits depends on the lepton flavours, and we lack of the limits for the $\tau$.
 Therefore, for each scenarios, we show two different figures, one for the case when $d_\gamma$ is universal, another one when $d_\gamma$ corresponds to $\tau$ flavour. Only the sensitivity at FASER-2 is shown here. FACET and MAPP-2 might also be potentially sensitive to the monophoton signature, while the detailed analyses to accounting the background and reconstruction efficiency are not provided yet in the literature, we only estimate the number of signal events of them in App.~\ref{App:futureLHC}.
 The current limits are taken from Ref.~\cite{Magill:2018jla, Jodlowski:2020vhr} considering the CHARM-II~\cite{CHARM-II:1989srx},
LSND~\cite{LSND:1996ubh}, MineBooNE~\cite{MiniBooNE:2007uho}, NOMAD~\cite{Vannucci:2014wna,NOMAD:1997pcg,NOMAD:1998pxi}, LEP~
\cite{OPAL:1994kgw, L3:1997exg}
, ATLAS and CMS at the LHC~\cite{ATLAS:2017nga,CMS:2015loa}~\footnote{The limits from CMS/ATLAS are updated using new analyses~\cite{ATLAS:2020uiq,CMS:2018fon}.} and Supernova SN 1987~\cite{Kamiokande-II:1987idp,Alekseev:1988gp,Bionta:1987qt} experiments.

\begin{figure}
\centering
\includegraphics[width=0.49\textwidth]{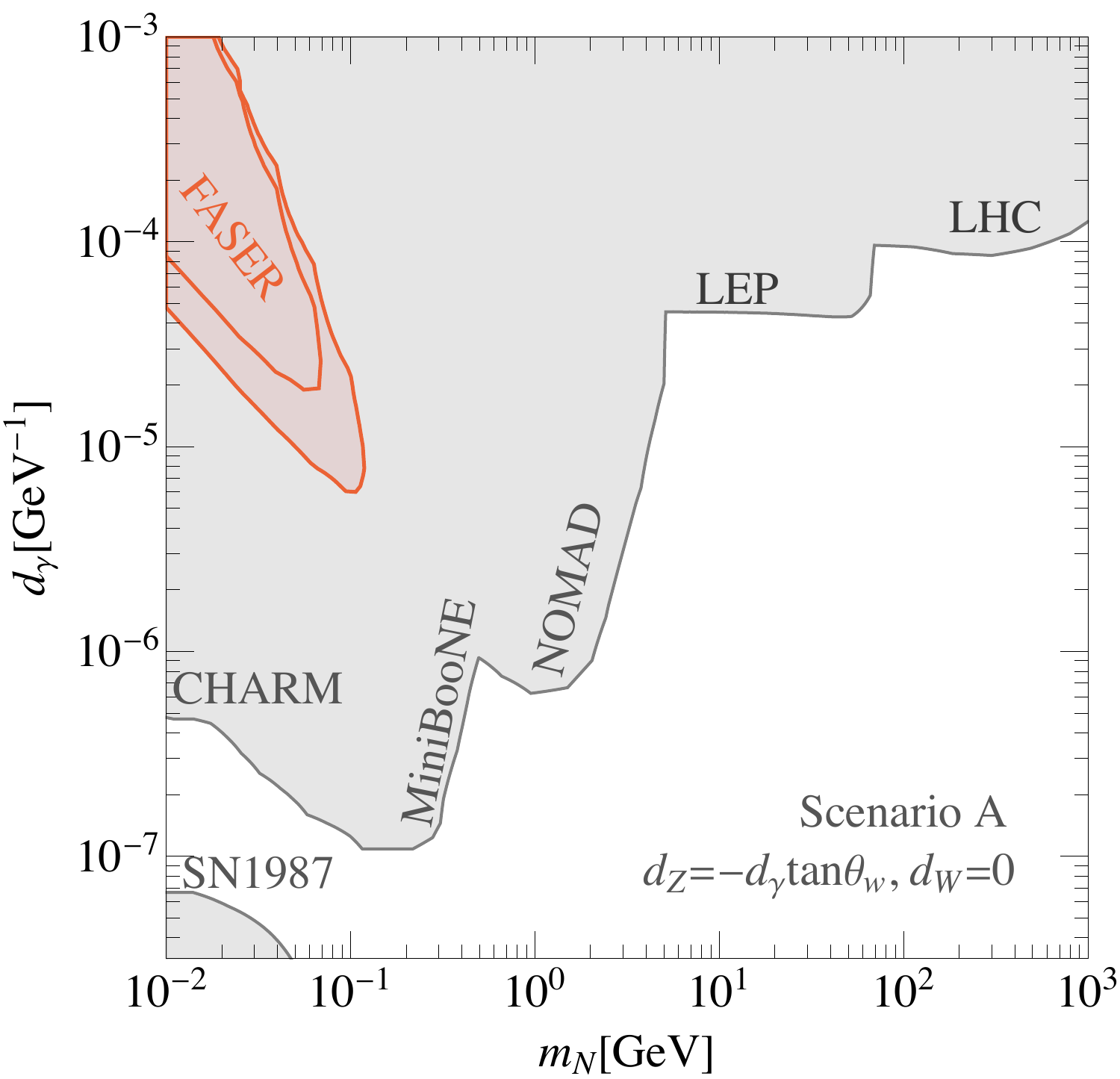}
\includegraphics[width=0.49\textwidth]{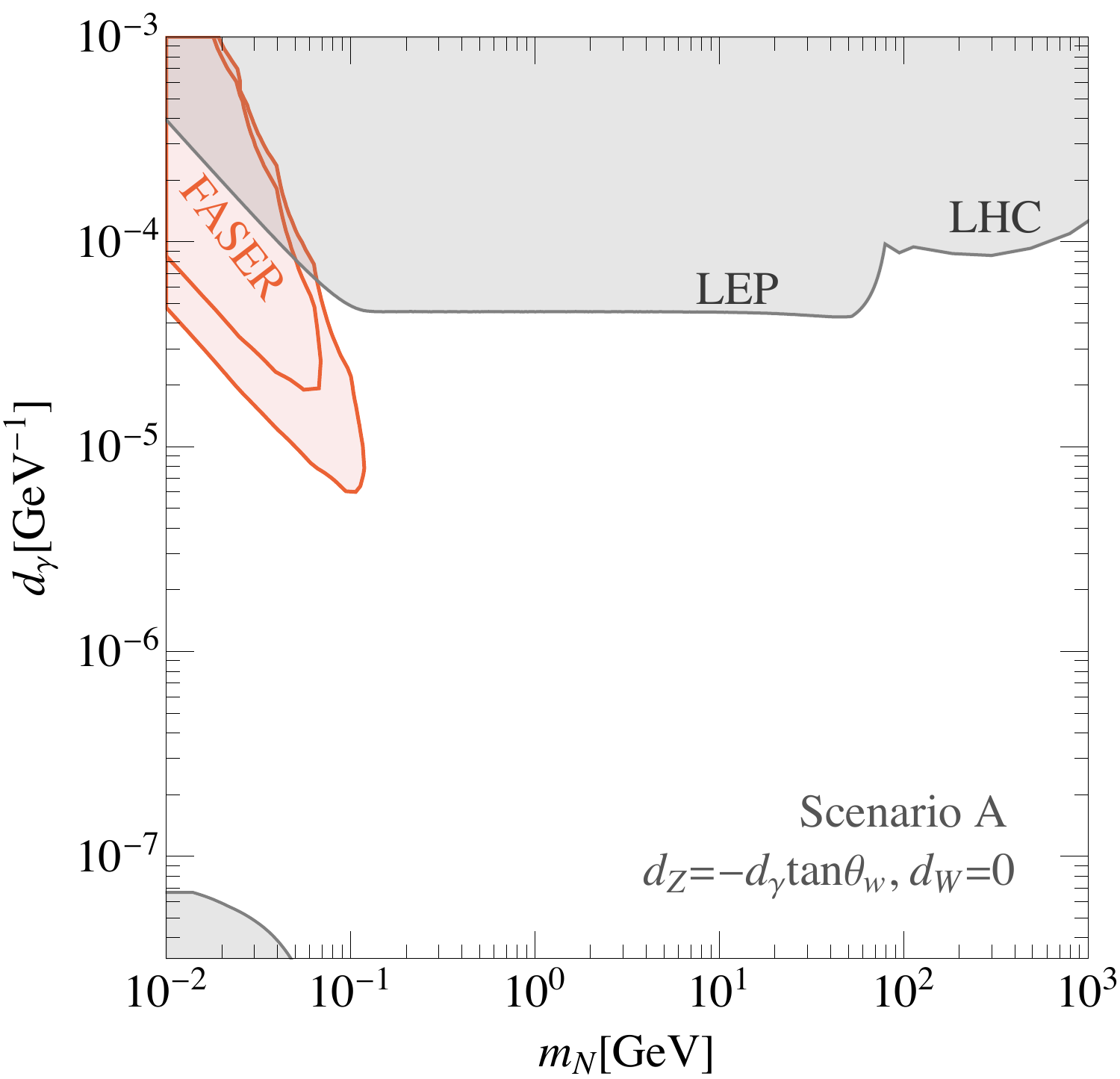}
\includegraphics[width=0.49\textwidth]{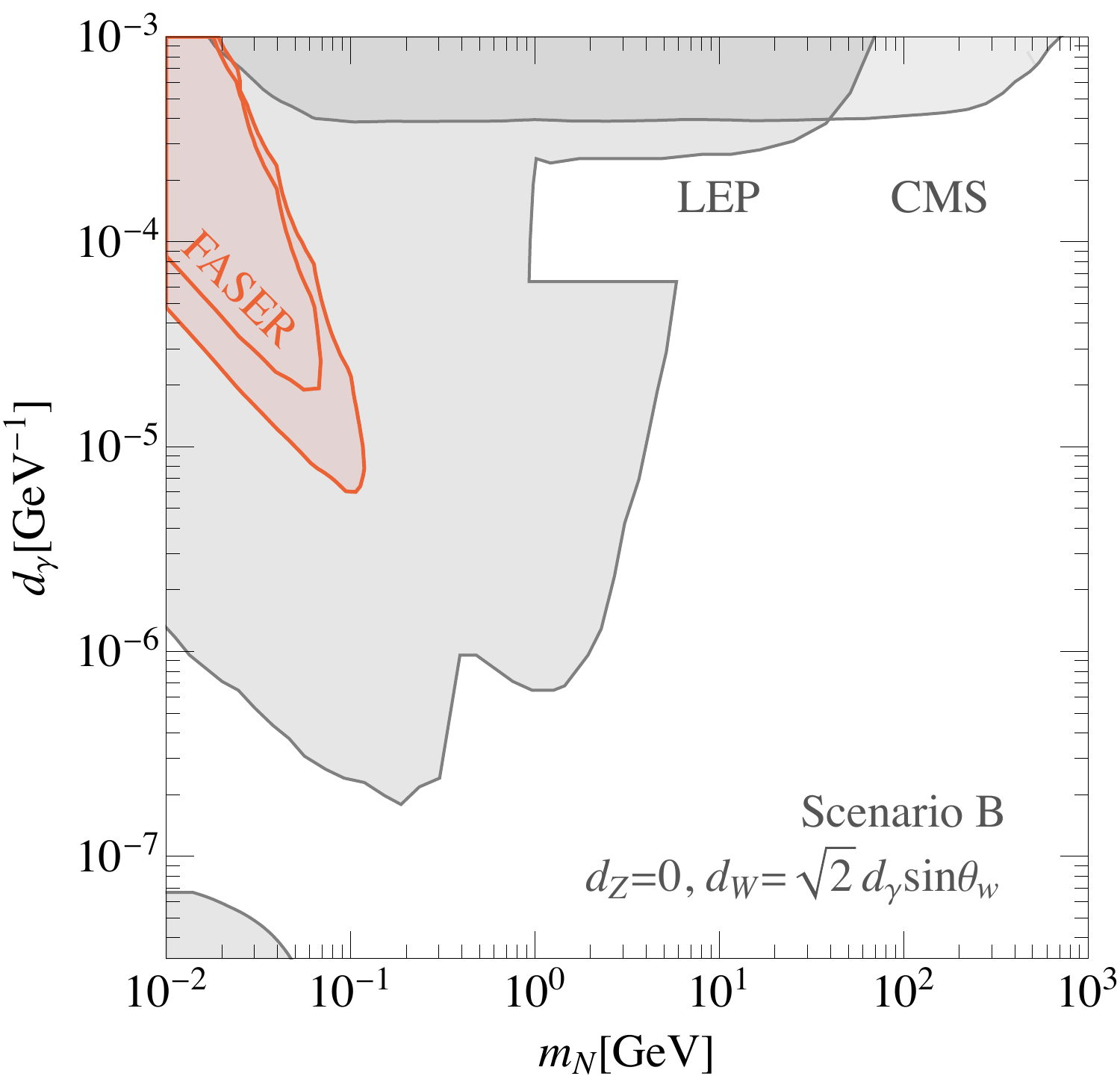}
\includegraphics[width=0.49\textwidth]{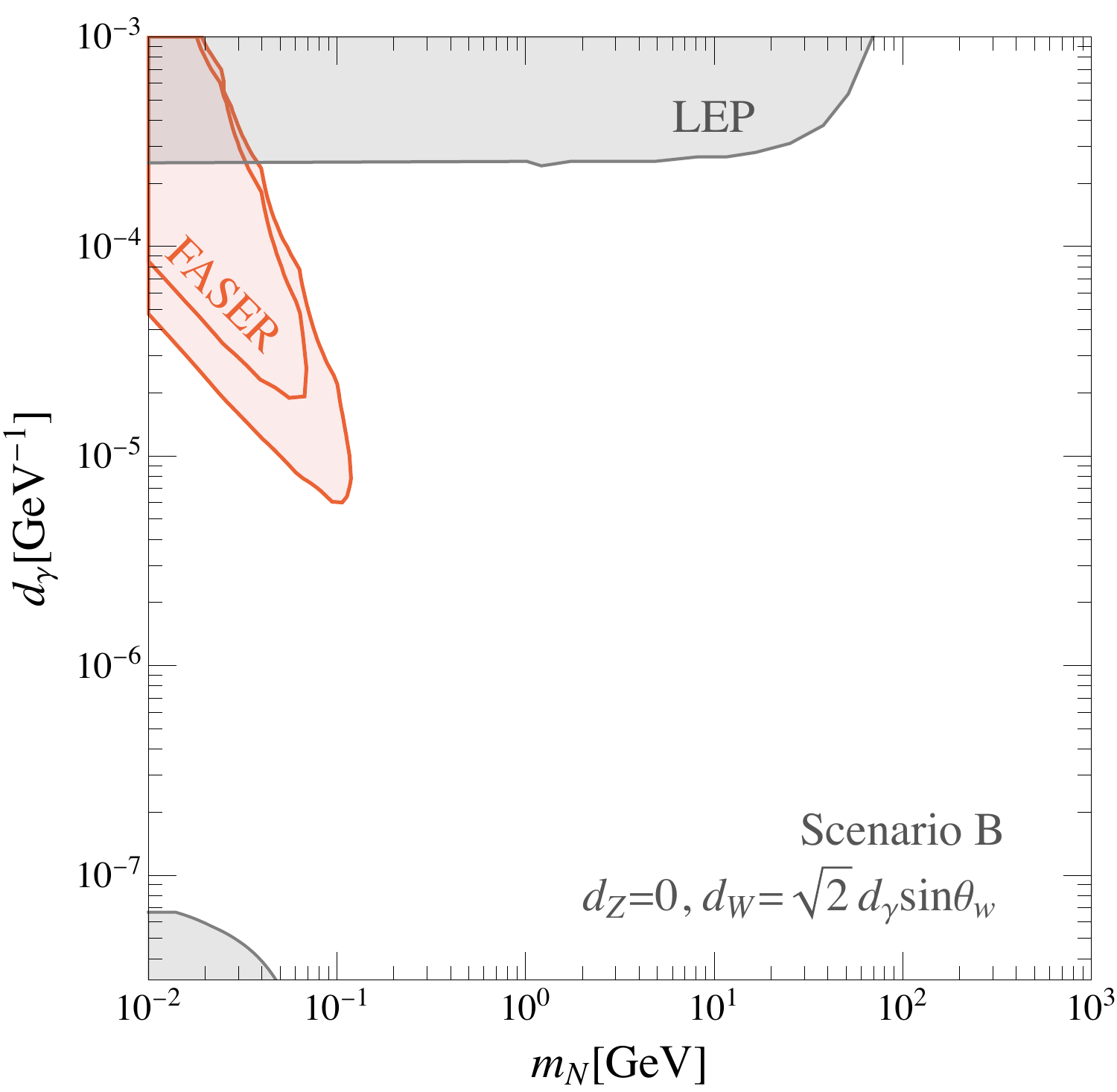}
\caption{Number of signal events of the LLP detectors including the FASER-2~(red) at the HL-LHC, in the ($m_N$, $d_\gamma$) plane for the Scenario A~(top) and B~(bottom). The red solid curve represents $N_{\text{signal}} = 3, 30$ at FASER-2 from bottom to up.
Current limits taken from Ref.~\cite{Magill:2018jla, Jodlowski:2020vhr} are overlaid for comparison. 
Left: For the universal coupling case. Right: For the case where the dipole portal couples to $\tau$ only.}
\label{fig:fdgA}
\end{figure}

In general, the LLP and other detectors at colliders are complementary to each other, as the LLP detectors probe where the $N$ is light, and CMS, ATLAS and LEP the opposite.
The results for Scenario A is demonstrated in Fig.~\ref{fig:fdgA}. 
For the universal coupling case as displayed in Fig.~\ref{fig:fdgA}~left, the curves for FASER-2 roughly tracks the curves where $L_N^{0} \sim \mathcal{O}(\text{m})$ as shown in Fig.~\ref{fig:fdecay}, until the coupling $d_\gamma \lesssim 10^{-5}$,
becoming too small to yield sufficient cross section for $m_N \gtrsim 10^{-1}$ GeV. FASER-2 can get where $d_\gamma \approx 10^{-5}$. The reason is already explained in Fig.~\ref{fig:feff}. In Fig.~\ref{fig:fdgA}~left, the results are shown in comparison with the current limits for the universal coupling case.
The coverage of the FASER-2 detectors in $m_N$ is within the ones of the CHARM experiment and neutrino scattering experiments, LSND~\cite{LSND:1996ubh} and MiniBooNE~\cite{MiniBooNE:2007uho}. Due to the enormous number of events using by these experiments, they have very high precision, therefore reaching lower $d_\gamma$ comparing to the FASER-2 detectors. Anyway, our efforts are not in vain, when we consider the case where the dipole portal couples to $\tau$ only in Fig.~\ref{fig:fdgA}~right. Now only the limits from the LEP, ATLAS and SN 1987 are effective, excluding $d_\gamma \gtrsim 10^{-4}$. Therefore, our results from the FASER-2 detectors are proved to be fairly useful, since they exceed the current limits by roughly one magnitude, when $m_N \lesssim 0.1$~GeV.

Now we move to the Scenario B, comparing to A, the FASER-2 has similar sensitivity, as their production cross section and decays branching ratio alike. Since $d_Z=0$ instead of $d_W=0$, therefore the current limits from ATLAS and LEP via $Z$ decays are no longer valid. The searches for $W$ mediated processes at the CMS applies, if the couplings are not $\tau$ only, since the searches aimed at light lepton final states. The searches for mono-photon signatures at the LEP are still applicable with much weaker limits. 
Thus, now the FASER-2 can give about two magnitude better sensitivity in the $\tau$ couplings only case.

\begin{figure}
\centering
\includegraphics[width=0.49\textwidth]{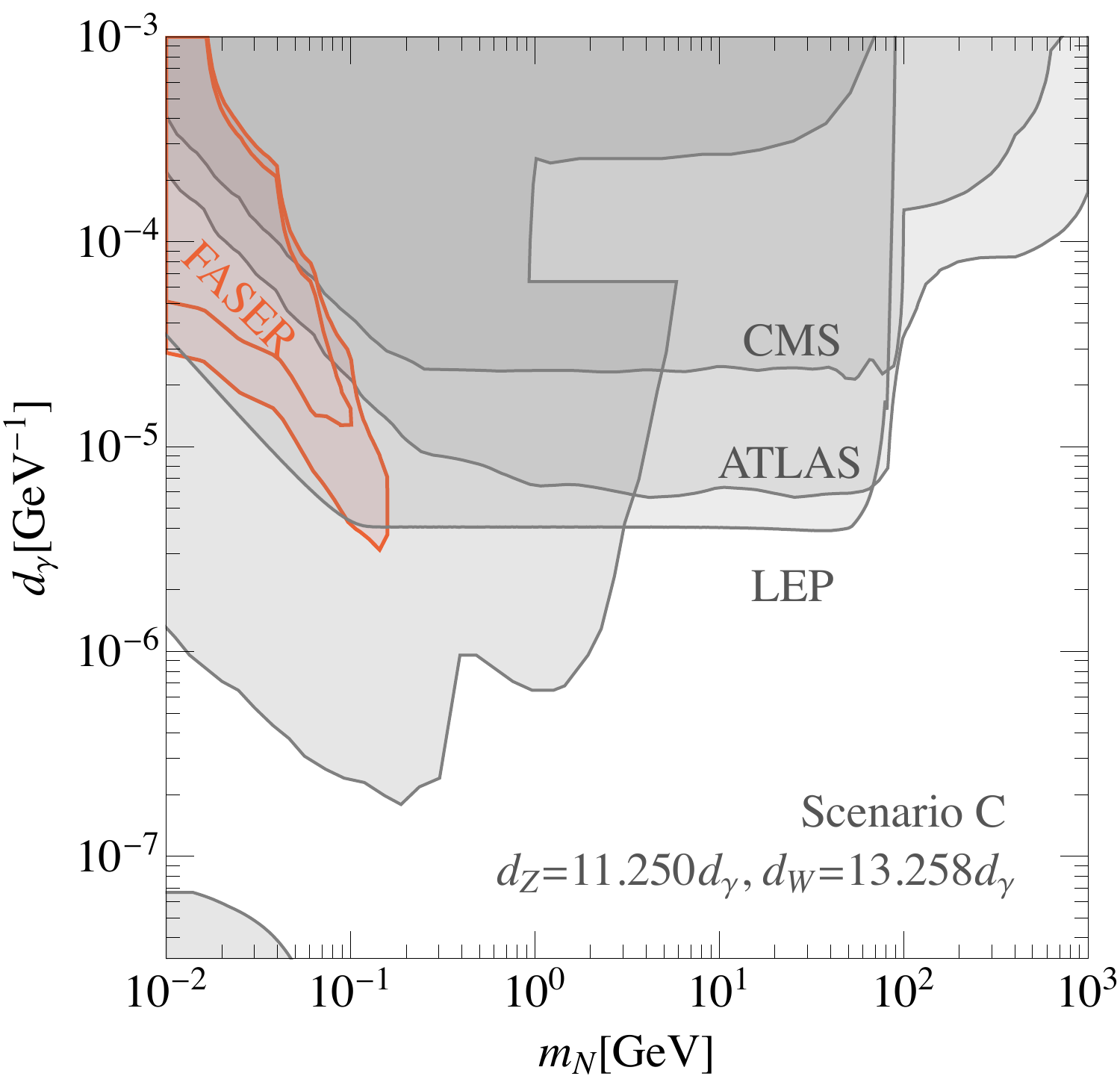}
\includegraphics[width=0.49\textwidth]{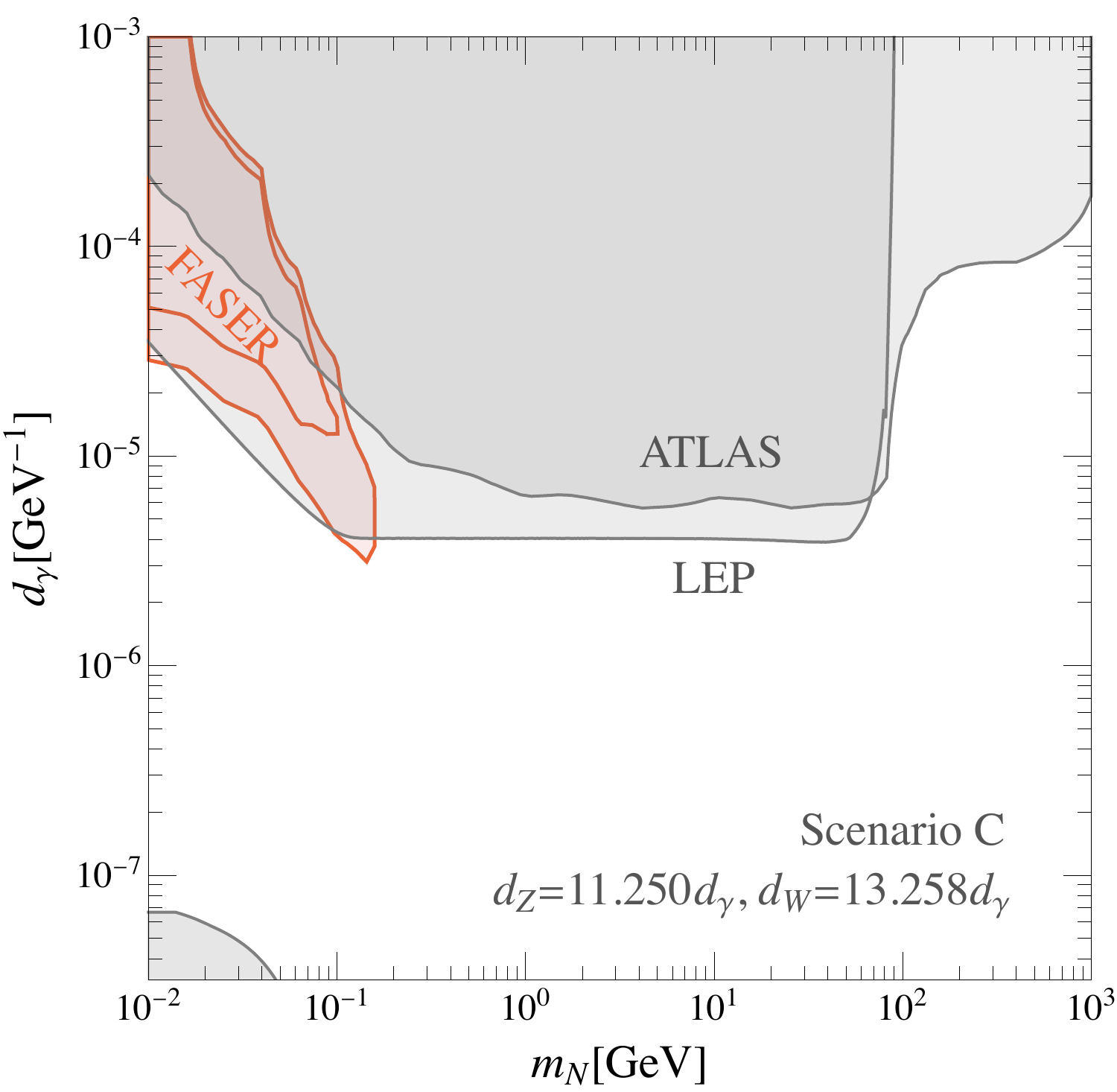}
\includegraphics[width=0.49\textwidth]{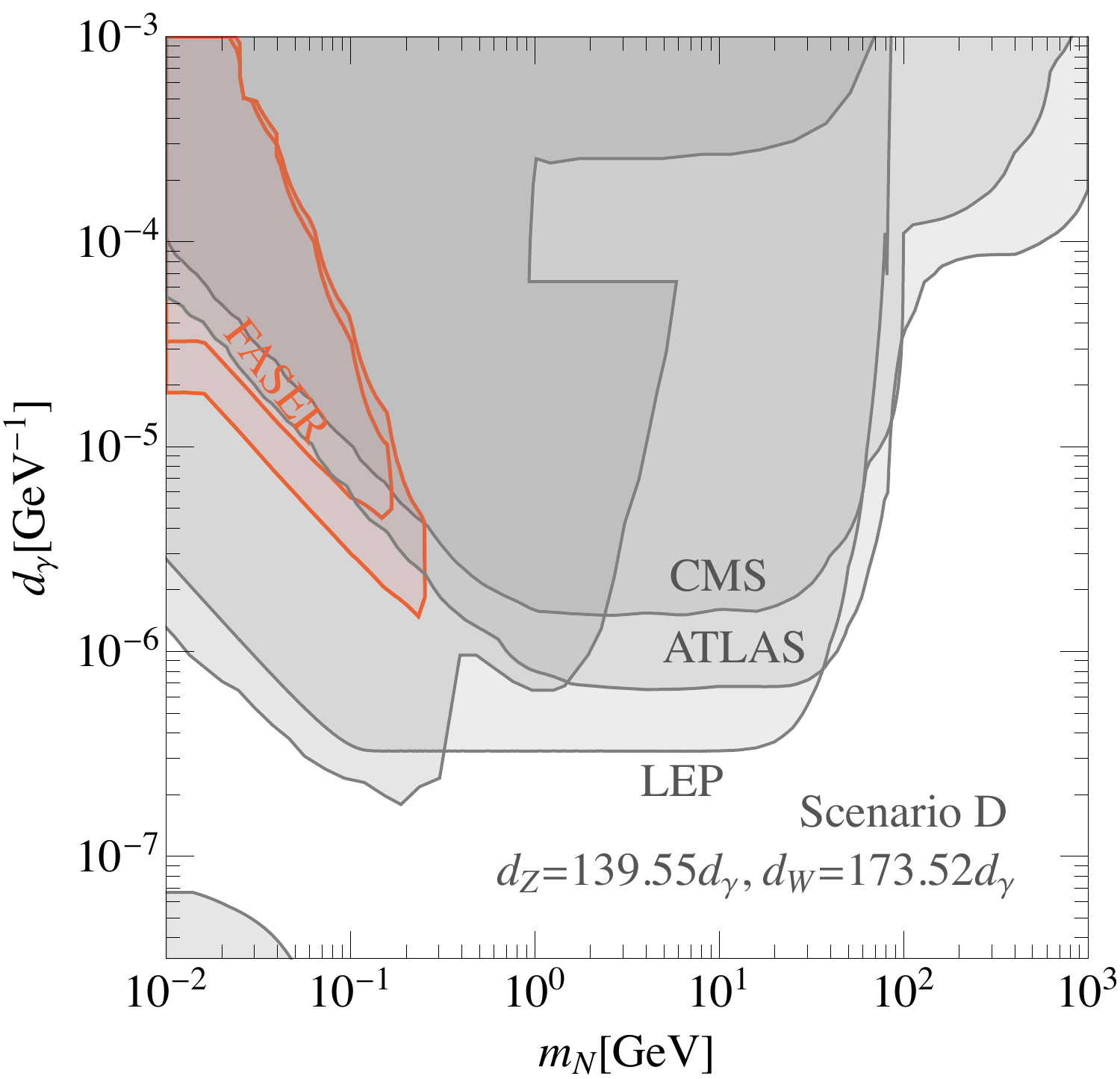}
\includegraphics[width=0.49\textwidth]{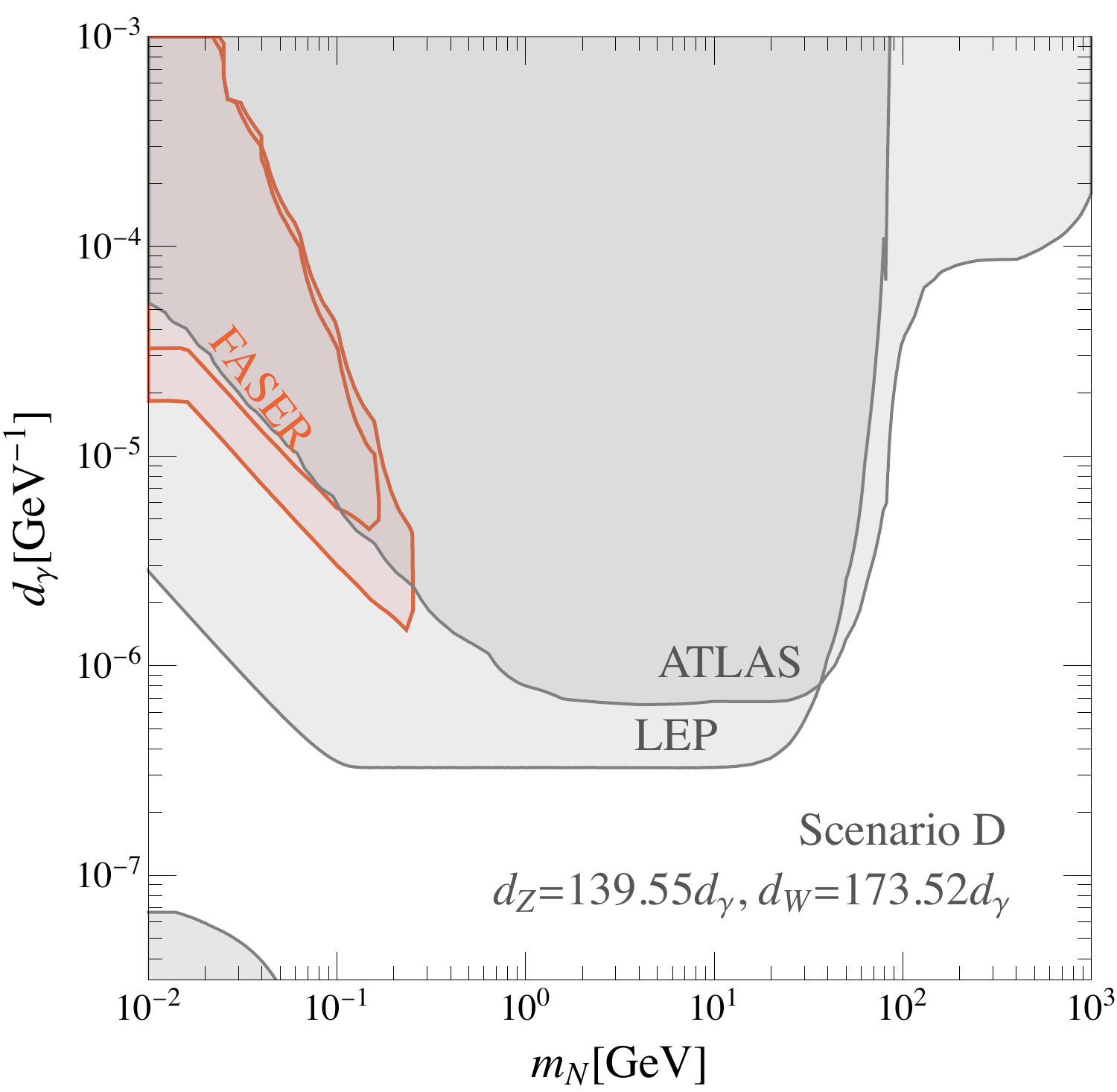}
\caption{Same as Fig.~\ref{fig:fdgA}, but for Scenario C~(top) and D~(bottom). The original limits from LEP, CMS and ATLAS are scaled, and hence shown prominently.}
\label{fig:fdgD}
\end{figure}

As for Scenario C and D, the high scale couplings $d_{W,Z}$ are effective. Since these couplings are about much {larger than $d_\gamma$} as indicated in Table~\ref{tab:scen}, the cross section {of $N$ production at LHC} is more than $10^{2,4}$ times larger {the one in Scenario A}. The larger cross section subsequently results in better reach at $d_\gamma$. From Fig.~\ref{fig:fdgD}, the lowest $d_\gamma$ can be probed is $10^{-5.5~(-6)}$ for FASER-2 in Scenario C~(D). Additionally, we redraw the current limits at high energy environment via analyses for prompt final states. We adopt ATLAS and CMS analyses, as well as the LEP analysis. Now these analyses benefited from the enlarged cross section as well, reaches to $d_\gamma \approx 10^{-(4.5-5.5)}$ in Scenario C, and $10^{-(5.5-6.5)}$ in Scenario D, only when $m_N \sim 0.1-90$~GeV. This is because these analyses is only sensitive to the HNL with $L_N^{\text{lab}} \lesssim 1$~m~\cite{ATLAS:2020uiq, CMS:2018fon, Zhang:2023nxy}, {and ${\rm Br}(N \rightarrow \nu \gamma)$ drops sharply once $m_N > M_{W,Z}$ as shown in Fig.~\ref{fig:fdecay}~left.

We compare them with the current limits, finding that the FASER-2 detector still can not compete with the low energy neutrino scattering and the CHARM experiments in the universal coupling case.  
When look at the case where the dipole portal couples to $\tau$ only in Fig.~\ref{fig:fdgD}~right, the low energy neutrino scattering and the CHARM as well as CMS experiments do not apply, as it is sensitive to the $e, \mu$ final states only. In Scenario C, now the FASER-2 yield similar sensitivity to the ones from LEP, and better than the ones from ATLAS. In Scenario D, it seems LEP and ATLAS fully take the advantange of large $d_{W,Z}$, leading to roughly half magnitude better limits.

\section{Conclusion}
\label{sec:con}
In pursuit of the explanation for the observed neutrino masses, many models assuming the existence of the HNLs are brought up. Among them, we focus on the neutrino dipole models within a dimension-6 EFT framework. This model contains high scale operators containing the couplings $d_{W,Z}$, which control the production of the HNLs at a high energy environment, e.g. the LHC. 

The current constraints are stringent on such models, with the upper limits $d_\gamma \sim 10^{-6}$ for $m_N < 1$~GeV, have already brought us to where the HNLs are long-lived. Although this case is already considered in Ref.~\cite{Jodlowski:2020vhr}, which employ the FASER-2 detector to search for the HNLs produced secondarily in neutrino interactions at the FASER$\nu$, and can probe lower $d_\gamma$ due to the large number of HNL produced from the neutrino interactions in the tungsten layers. 
The dependence on the high scale operators $d_{W,Z}$ is however not considered.
In this paper, we discuss the effects of different relations between $d_{W,Z}$, and the low scale coupling $d_\gamma$, then estimate the sensitivity of the LLP detector, FASER-2, with the HNL produced primarily. 

The LLP detectors, located far away from the IP of the LHC, can be senstive to new particles which are light and weak coupled to the SM, leading to long decay length. Although weak couplings can lead to low statistics, this is overcome since the high scale couplings can produce large number of the HNLs, no matter the low scale decay coupling is.

We choose four scenarios for comparison to show the dependence on the relations between $d_{W,Z}$ and 
$d_\gamma$. {In Scenarios A and B with either $d_{W}=0$ or $d_{Z=0}$ and $d_{Z/W}$ is comparable to $d_\gamma$, the production rates are mainly controlled by the $d_\gamma$, while Scenario C and D dominantly controlled by $d_{W,Z}$ since $d_{W}$ and $d_{Z}$ are far larger than  $d_\gamma$.}
For the former scenarios A and B, we show that the FASER-2 detectors can reach $d_\gamma \approx 10^{-5}$ when $m_N \lesssim 0.1$~GeV. Although this parameter space is already ruled out by neutrino scattering experiments, e.g. MiniBooNE and LSND, as well as the CHARM experiment, for $d_\gamma$ corresponds to the $e, \mu$ flavours or if it is universal, it is
about one or two magnitude lower than the current limits including the ones at LEP, CMS and ATLAS, when the dipole only couples to $\tau$. For the latter scenarios C and D, since the production is enhanced by the choices of $d_{W,Z}$, the FASER-2 detectors can now reach $d_\gamma \approx 10^{-6}$. However, since the productions at LEP, CMS and ATLAS are directly connected to the $d_{W,Z}$, now the limits from them is comparable to the FASER-2 in Scenario C, and better for half magnitude in Scenario D.

We also shown the projected number of signal events for the proposed MAPP-2 and FACET detectors in App.~\ref{App:futureLHC}, which can potentially yield better sensitivity if the background can be controlled, and we leave the dedicated analyses for future study.

\acknowledgments
We thank Zeren Simon Wang and Arsenii Titov for useful discussions.
WL is supported by National Natural Science Foundation of China (Grant No.12205153), and the 2021 Jiangsu Shuangchuang (Mass Innovation and Entrepreneurship) Talent Program (JSSCBS20210213). YZ is supported in part by the National Natural Science Foundation of China (Grant No. 11805001) and the Fundamental Research Funds for the Central Universities (Grant No. JZ2023HGTB0222).
\appendix
\section{Projected sensitivity of other future LHC experiments}
\label{App:futureLHC}
\begin{figure}
\centering
\includegraphics[width=0.49\textwidth]{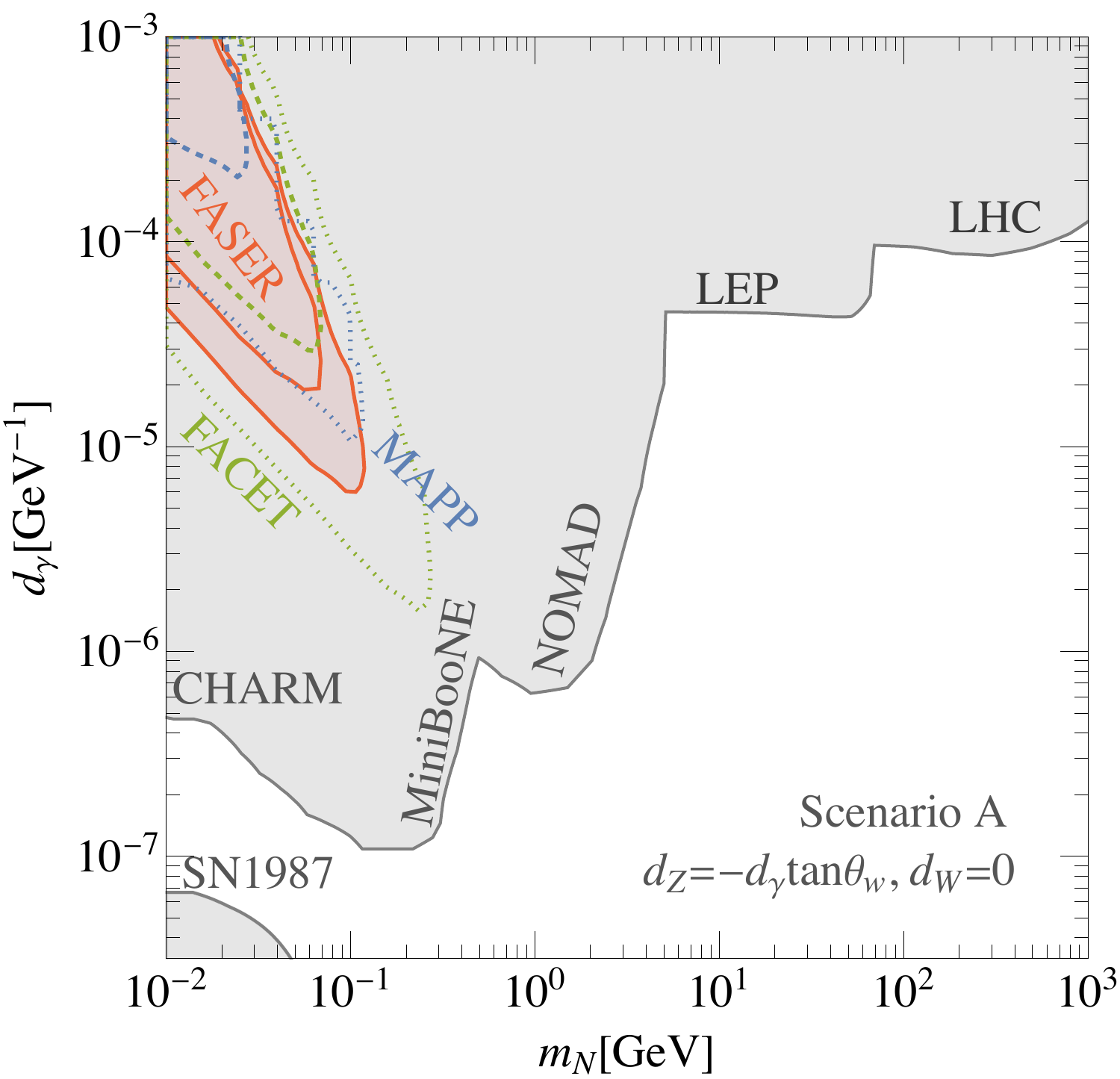}
\includegraphics[width=0.49\textwidth]{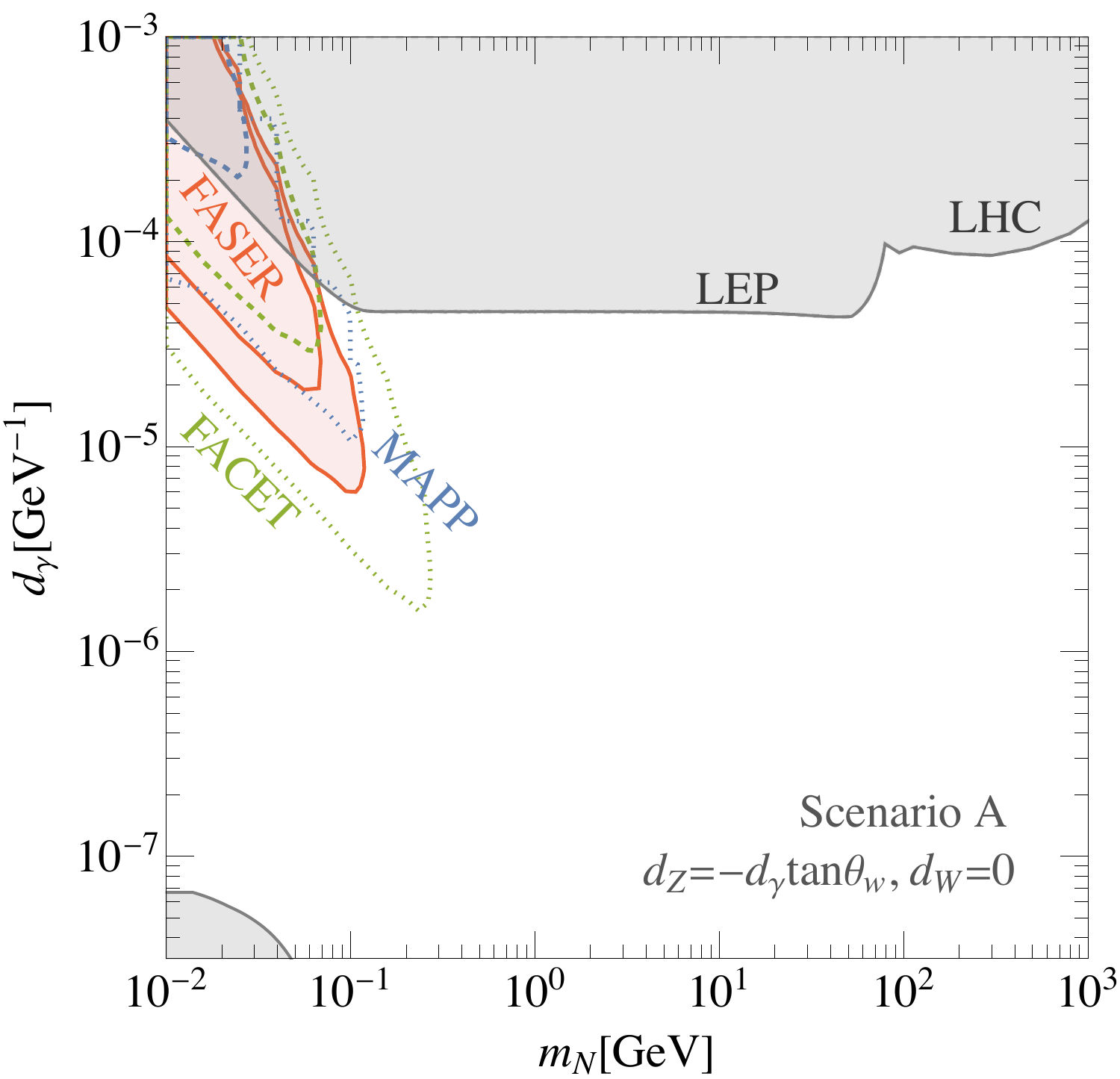}
\includegraphics[width=0.49\textwidth]{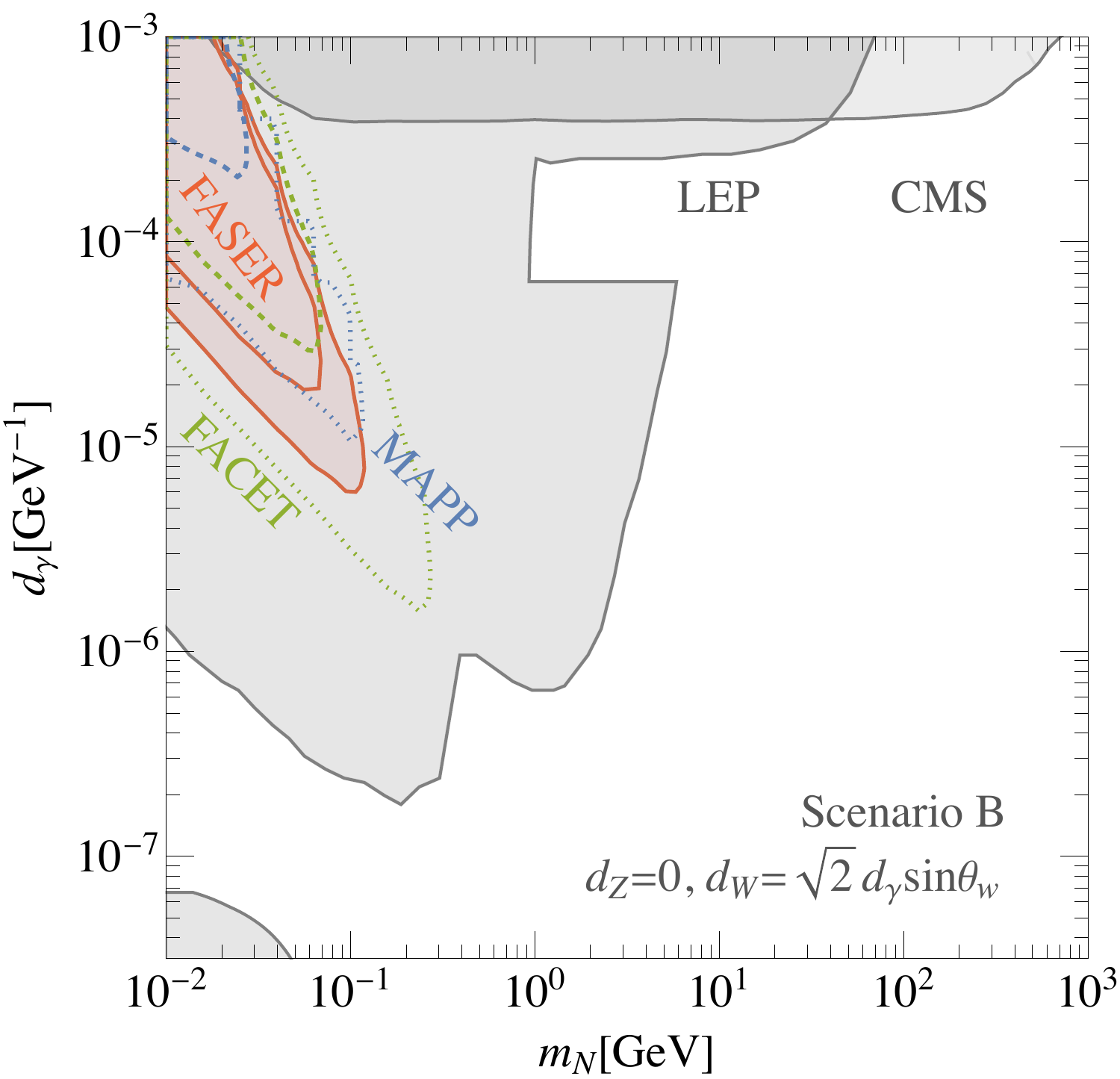}
\includegraphics[width=0.49\textwidth]{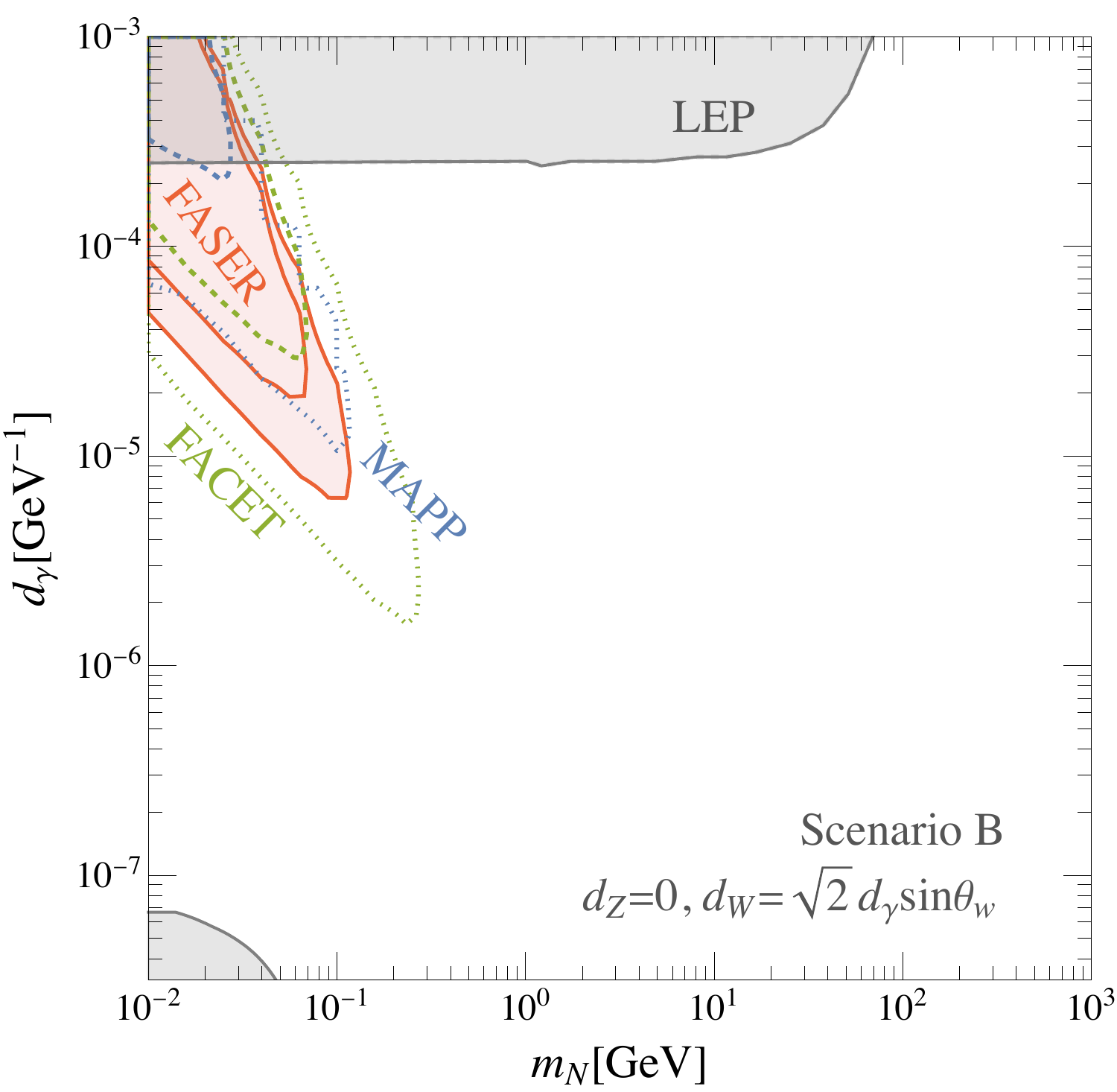}
\caption{Number of signal events of the LLP detectors including the FASER-2~(red), MAPP-2~(blue) and FACET~(green) at the HL-LHC, in the ($m_N$, $d_\gamma$) plane for the Scenario A~(top) and B~(bottom). The dotted and dashed curves corresponds to $N_{\text{signal}} = 3$, 1000 at MAPP~(blue) and FACET~(green).
Current limits taken from Ref.~\cite{Magill:2018jla, Jodlowski:2020vhr} are overlaid for comparison. 
Left: For the universal coupling case. Right: For the case where the dipole portal couples to $\tau$ only.}
\label{fig:fdgAapp}
\end{figure}

\begin{figure}
\centering
\includegraphics[width=0.49\textwidth]{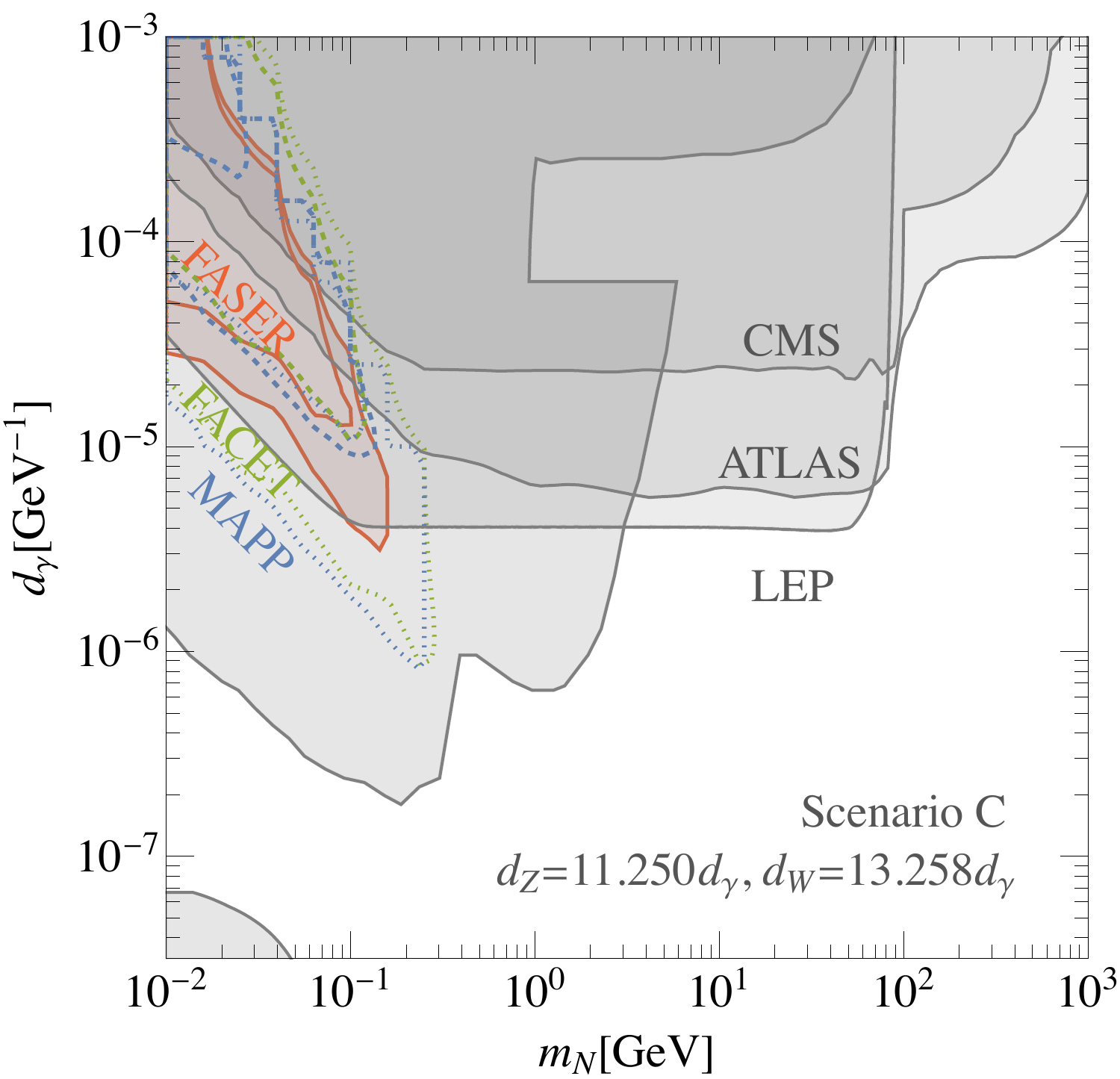}
\includegraphics[width=0.49\textwidth]{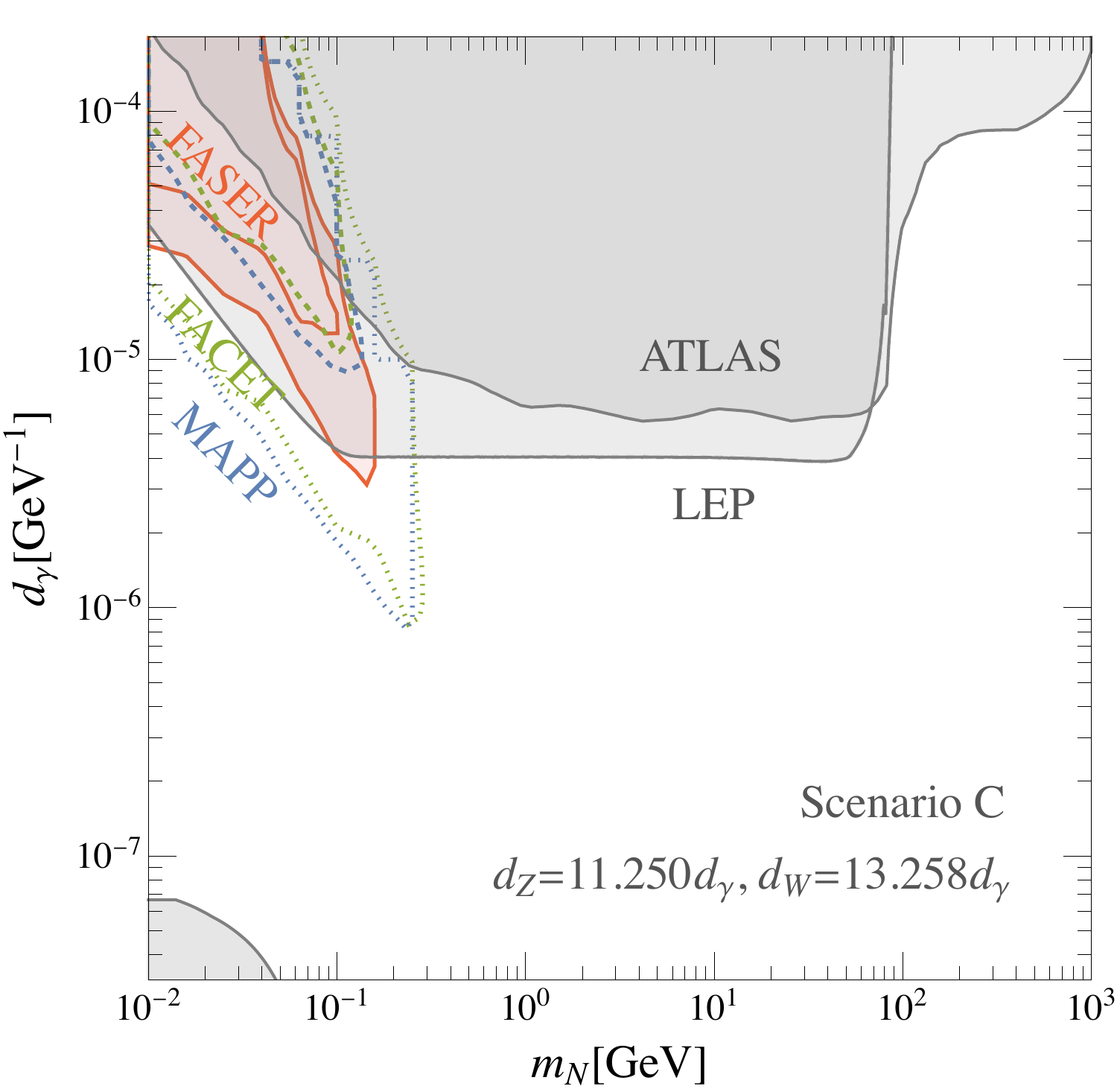}
\includegraphics[width=0.49\textwidth]{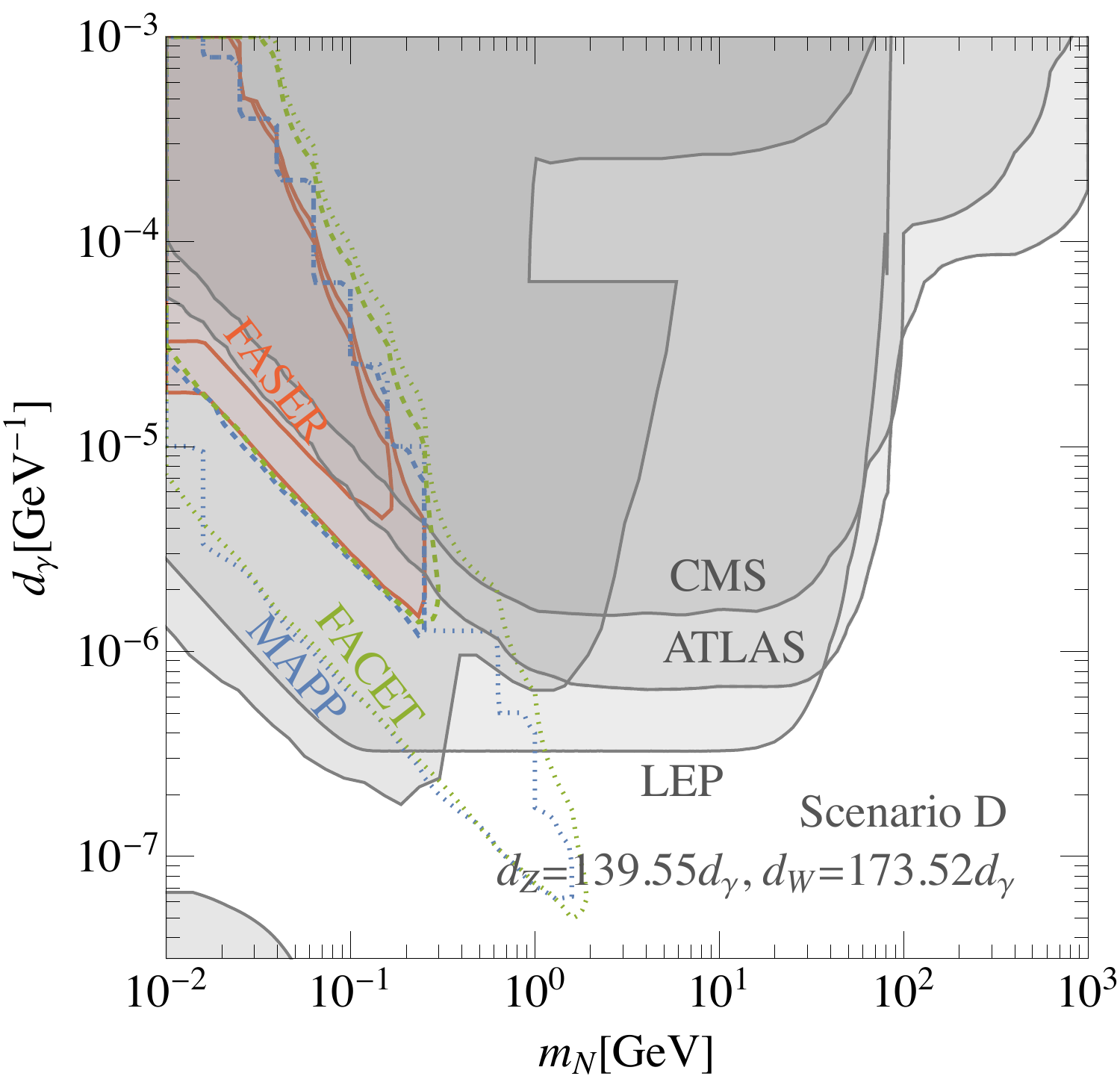}
\includegraphics[width=0.49\textwidth]{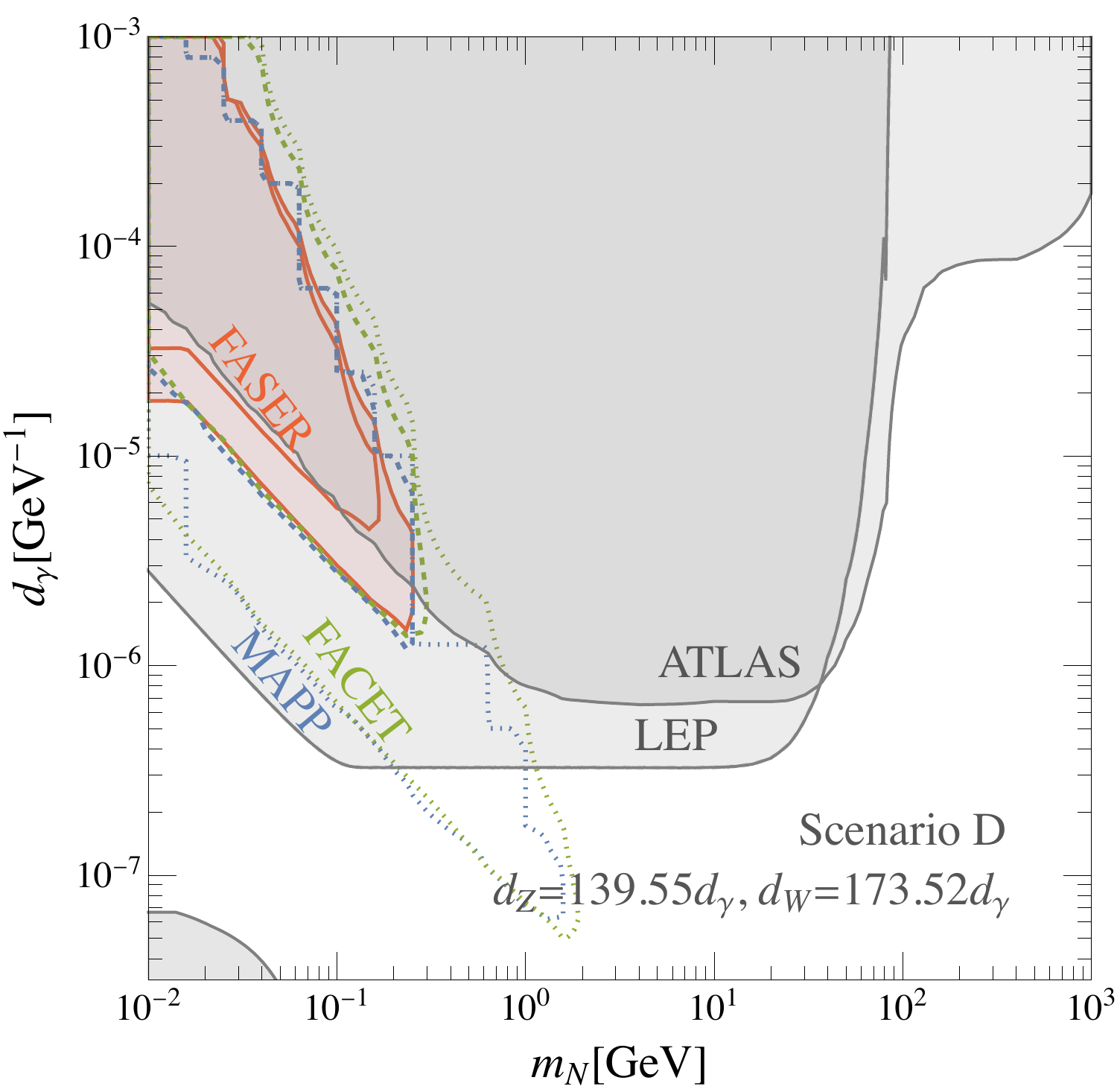}
\caption{Same as Fig.~\ref{fig:fdgAapp}, but for Scenario C~(top) and D~(bottom).}
\label{fig:fdgDapp}
\end{figure}
The FACET and MAPP-2 can potentially probe the diopole couplings as well. Since the detailed discussion of the background is not provided yet in literature, we only show the fixed number of signal events, and employ the same kinematic trigger as FASER-2. $N_{\text{signal}}=3,~1000$ is taken for MAPP-2 and FACET, to roughly indicate the potential reach of the model, if the number of background events can be controlled in a certain amount. 

Their results are shown in Fig.~\ref{fig:fdgAapp} and~\ref{fig:fdgDapp}. In Scenarios A and B, FACET can yield roughly half magnitude better sensitivity than the FASER-2, and even more than the MAPP-2, leading to fairly positive prospect epecially in the $\tau$ couplings only case, if the background can be controlled. When it comes to the Scenario C and D, MAPP and FACET now have very similar sensitivity, and can potentially exceed all the current limits by about half magnitude when $m_N \sim 1$ GeV, if the background can be controlled. In a pessimistic view, if the background events can not be controlled, positive prospect can still be achieved even if we ask for $10^3$ signal events, which corresponds to $10^6$ number of background events, the sensitivity is still comparable to the ones from FASER-2.

Comparing the LLP detectors, those located at the forward direction, including FASER-2 and FACET, have shown drastic different geometrical efficiencies in these two scenarios, due to the different contribution from the $\gamma$ mediated processes. For them, the meson mediated processes might proved to be more powerful, and we leave them in the upcoming work.
\bibliographystyle{JHEP}
\bibliography{submit.bib}

\providecommand{\href}[2]{#2}\begingroup\raggedright\begin{thebibliography}{10}

\bibitem{Balaji:2019fxd}
S.~Balaji, M.~Ramirez-Quezada and Y.-L. Zhou, \emph{{CP violation and circular
  polarisation in neutrino radiative decay}},
  \href{http://dx.doi.org/10.1007/JHEP04(2020)178}{\emph{JHEP} {\bf 04} (2020)
  178}, [\href{http://arxiv.org/abs/1910.08558}{{\tt 1910.08558}}].

\bibitem{Balaji:2020oig}
S.~Balaji, M.~Ramirez-Quezada and Y.-L. Zhou, \emph{{CP violation in neutral
  lepton transition dipole moment}},
  \href{http://dx.doi.org/10.1007/JHEP12(2020)090}{\emph{JHEP} {\bf 12} (2020)
  090}, [\href{http://arxiv.org/abs/2008.12795}{{\tt 2008.12795}}].

\bibitem{Delgado:2022fea}
F.~Delgado, L.~Duarte, J.~Jones-Perez, C.~Manrique-Chavil and S.~Pe\~na,
  \emph{{Assessment of the dimension-5 seesaw portal and impact of exotic Higgs
  decays on non-pointing photon searches}},
  \href{http://dx.doi.org/10.1007/JHEP09(2022)079}{\emph{JHEP} {\bf 09} (2022)
  079}, [\href{http://arxiv.org/abs/2205.13550}{{\tt 2205.13550}}].

\bibitem{Barducci:2022gdv}
D.~Barducci, E.~Bertuzzo, M.~Taoso and C.~Toni, \emph{{Probing right-handed
  neutrinos dipole operators}},  \href{http://arxiv.org/abs/2209.13469}{{\tt
  2209.13469}}.

\bibitem{Ding:2019tqq}
J.-N. Ding, Q.~Qin and F.-S. Yu, \emph{{Heavy neutrino searches at future
  $Z$-factories}},
  \href{http://dx.doi.org/10.1140/epjc/s10052-019-7277-3}{\emph{Eur. Phys. J.
  C} {\bf 79} (2019) 766}, [\href{http://arxiv.org/abs/1903.02570}{{\tt
  1903.02570}}].

\bibitem{Shen:2022ffi}
Y.-F. Shen, J.-N. Ding and Q.~Qin, \emph{{Monojet search for heavy neutrinos at
  future Z-factories}},
  \href{http://dx.doi.org/10.1140/epjc/s10052-022-10301-4}{\emph{Eur. Phys. J.
  C} {\bf 82} (2022) 398}, [\href{http://arxiv.org/abs/2201.05831}{{\tt
  2201.05831}}].

\bibitem{Deppisch:2018eth}
F.~F. Deppisch, W.~Liu and M.~Mitra, \emph{{Long-lived Heavy Neutrinos from
  Higgs Decays}}, \href{http://dx.doi.org/10.1007/JHEP08(2018)181}{\emph{JHEP}
  {\bf 08} (2018) 181}, [\href{http://arxiv.org/abs/1804.04075}{{\tt
  1804.04075}}].

\bibitem{Deppisch:2019kvs}
F.~Deppisch, S.~Kulkarni and W.~Liu, \emph{{Heavy neutrino production via $Z'$
  at the lifetime frontier}},
  \href{http://dx.doi.org/10.1103/PhysRevD.100.035005}{\emph{Phys. Rev. D} {\bf
  100} (2019) 035005}, [\href{http://arxiv.org/abs/1905.11889}{{\tt
  1905.11889}}].

\bibitem{Liu:2022kid}
W.~Liu, S.~Kulkarni and F.~F. Deppisch, \emph{{Heavy neutrinos at the FCC-hh in
  the U(1)B-L model}},
  \href{http://dx.doi.org/10.1103/PhysRevD.105.095043}{\emph{Phys. Rev. D} {\bf
  105} (2022) 095043}, [\href{http://arxiv.org/abs/2202.07310}{{\tt
  2202.07310}}].

\bibitem{Liu:2021akf}
W.~Liu, K.-P. Xie and Z.~Yi, \emph{{Testing leptogenesis at the LHC and future
  muon colliders: A Z' scenario}},
  \href{http://dx.doi.org/10.1103/PhysRevD.105.095034}{\emph{Phys. Rev. D} {\bf
  105} (2022) 095034}, [\href{http://arxiv.org/abs/2109.15087}{{\tt
  2109.15087}}].

\bibitem{Liu:2022ugx}
W.~Liu, J.~Li, J.~Li and H.~Sun, \emph{{Testing the seesaw mechanisms via
  displaced right-handed neutrinos from a light scalar at the HL-LHC}},
  \href{http://dx.doi.org/10.1103/PhysRevD.106.015019}{\emph{Phys. Rev. D} {\bf
  106} (2022) 015019}, [\href{http://arxiv.org/abs/2204.03819}{{\tt
  2204.03819}}].

\bibitem{Beltran:2022ast}
R.~Beltr\'an, G.~Cottin, J.~C. Helo, M.~Hirsch, A.~Titov and Z.~S. Wang,
  \emph{{Long-lived heavy neutral leptons from mesons in effective field
  theory}}, \href{http://dx.doi.org/10.1007/JHEP01(2023)015}{\emph{JHEP} {\bf
  01} (2023) 015}, [\href{http://arxiv.org/abs/2210.02461}{{\tt 2210.02461}}].

\bibitem{Zhou:2021ylt}
G.~Zhou, J.~Y. G\"unther, Z.~S. Wang, J.~de~Vries and H.~K. Dreiner,
  \emph{{Long-lived sterile neutrinos at Belle II in effective field theory}},
  \href{http://dx.doi.org/10.1007/JHEP04(2022)057}{\emph{JHEP} {\bf 04} (2022)
  057}, [\href{http://arxiv.org/abs/2111.04403}{{\tt 2111.04403}}].

\bibitem{Abada:2018sfh}
A.~Abada, N.~Bernal, M.~Losada and X.~Marcano, \emph{{Inclusive Displaced
  Vertex Searches for Heavy Neutral Leptons at the LHC}},
  \href{http://dx.doi.org/10.1007/JHEP01(2019)093}{\emph{JHEP} {\bf 01} (2019)
  093}, [\href{http://arxiv.org/abs/1807.10024}{{\tt 1807.10024}}].

\bibitem{Fernandez-Martinez:2022gsu}
E.~Fern\'andez-Mart\'\i{}nez, X.~Marcano and D.~Naredo-Tuero, \emph{{HNL mass
  degeneracy: implications for low-scale seesaws, LNV at colliders and
  leptogenesis}},  \href{http://arxiv.org/abs/2209.04461}{{\tt 2209.04461}}.

\bibitem{Abada:2022wvh}
A.~Abada, P.~Escribano, X.~Marcano and G.~Piazza, \emph{{Collider searches for
  heavy neutral leptons: beyond simplified scenarios}},
  \href{http://dx.doi.org/10.1140/epjc/s10052-022-11011-7}{\emph{Eur. Phys. J.
  C} {\bf 82} (2022) 1030}, [\href{http://arxiv.org/abs/2208.13882}{{\tt
  2208.13882}}].

\bibitem{Arganda:2015ija}
E.~Arganda, M.~J. Herrero, X.~Marcano and C.~Weiland, \emph{{Exotic
  \ensuremath{\mu}\ensuremath{\tau}jj events from heavy ISS neutrinos at the
  LHC}}, \href{http://dx.doi.org/10.1016/j.physletb.2015.11.013}{\emph{Phys.
  Lett. B} {\bf 752} (2016) 46--50},
  [\href{http://arxiv.org/abs/1508.05074}{{\tt 1508.05074}}].

\bibitem{Bai:2022lbv}
L.~Bai, Y.-n. Mao and K.~Wang, \emph{{Probe the Mixing Parameter $|V_{\tau
  N}|^2$ for Heavy Neutrinos}},  \href{http://arxiv.org/abs/2211.00309}{{\tt
  2211.00309}}.

\bibitem{Das:2017nvm}
A.~Das and N.~Okada, \emph{{Bounds on heavy Majorana neutrinos in type-I seesaw
  and implications for collider searches}},
  \href{http://dx.doi.org/10.1016/j.physletb.2017.09.042}{\emph{Phys. Lett. B}
  {\bf 774} (2017) 32--40}, [\href{http://arxiv.org/abs/1702.04668}{{\tt
  1702.04668}}].

\bibitem{Das:2012ze}
A.~Das and N.~Okada, \emph{{Inverse seesaw neutrino signatures at the LHC and
  ILC}}, \href{http://dx.doi.org/10.1103/PhysRevD.88.113001}{\emph{Phys. Rev.
  D} {\bf 88} (2013) 113001}, [\href{http://arxiv.org/abs/1207.3734}{{\tt
  1207.3734}}].

\bibitem{Das:2015toa}
A.~Das and N.~Okada, \emph{{Improved bounds on the heavy neutrino productions
  at the LHC}}, \href{http://dx.doi.org/10.1103/PhysRevD.93.033003}{\emph{Phys.
  Rev. D} {\bf 93} (2016) 033003}, [\href{http://arxiv.org/abs/1510.04790}{{\tt
  1510.04790}}].

\bibitem{Das:2016hof}
A.~Das, P.~Konar and S.~Majhi, \emph{{Production of Heavy neutrino in
  next-to-leading order QCD at the LHC and beyond}},
  \href{http://dx.doi.org/10.1007/JHEP06(2016)019}{\emph{JHEP} {\bf 06} (2016)
  019}, [\href{http://arxiv.org/abs/1604.00608}{{\tt 1604.00608}}].

\bibitem{Alok:2022pdn}
A.~K. Alok, N.~R. Singh~Chundawat and A.~Mandal, \emph{{Cosmic neutrino flux
  and spin flavor oscillations in intergalactic medium}},
  \href{http://dx.doi.org/10.1016/j.physletb.2023.137791}{\emph{Phys. Lett. B}
  {\bf 839} (2023) 137791}, [\href{http://arxiv.org/abs/2207.13034}{{\tt
  2207.13034}}].

\bibitem{Butterworth:2019iff}
J.~M. Butterworth, M.~Chala, C.~Englert, M.~Spannowsky and A.~Titov,
  \emph{{Higgs phenomenology as a probe of sterile neutrinos}},
  \href{http://dx.doi.org/10.1103/PhysRevD.100.115019}{\emph{Phys. Rev. D} {\bf
  100} (2019) 115019}, [\href{http://arxiv.org/abs/1909.04665}{{\tt
  1909.04665}}].

\bibitem{Abdullahi:2022jlv}
A.~M. Abdullahi et~al., \emph{{The Present and Future Status of Heavy Neutral
  Leptons}},  in \emph{{2022 Snowmass Summer Study}}, 3, 2022.
\newblock \href{http://arxiv.org/abs/2203.08039}{{\tt 2203.08039}}.

\bibitem{Amrith:2018yfb}
S.~Amrith, J.~M. Butterworth, F.~F. Deppisch, W.~Liu, A.~Varma and D.~Yallup,
  \emph{{LHC Constraints on a $B-L$ Gauge Model using Contur}},
  \href{http://dx.doi.org/10.1007/JHEP05(2019)154}{\emph{JHEP} {\bf 05} (2019)
  154}, [\href{http://arxiv.org/abs/1811.11452}{{\tt 1811.11452}}].

\bibitem{Magill:2018jla}
G.~Magill, R.~Plestid, M.~Pospelov and Y.-D. Tsai, \emph{{Dipole Portal to
  Heavy Neutral Leptons}},
  \href{http://dx.doi.org/10.1103/PhysRevD.98.115015}{\emph{Phys. Rev. D} {\bf
  98} (2018) 115015}, [\href{http://arxiv.org/abs/1803.03262}{{\tt
  1803.03262}}].

\bibitem{Aparici:2009fh}
A.~Aparici, K.~Kim, A.~Santamaria and J.~Wudka, \emph{{Right-handed neutrino
  magnetic moments}},
  \href{http://dx.doi.org/10.1103/PhysRevD.80.013010}{\emph{Phys. Rev. D} {\bf
  80} (2009) 013010}, [\href{http://arxiv.org/abs/0904.3244}{{\tt 0904.3244}}].

\bibitem{Giunti:2014ixa}
C.~Giunti and A.~Studenikin, \emph{{Neutrino electromagnetic interactions: a
  window to new physics}},
  \href{http://dx.doi.org/10.1103/RevModPhys.87.531}{\emph{Rev. Mod. Phys.}
  {\bf 87} (2015) 531}, [\href{http://arxiv.org/abs/1403.6344}{{\tt
  1403.6344}}].

\bibitem{Aparici:2013xga}
A.~Aparici, \emph{{Exotic properties of neutrinos using effective Lagrangians
  and specific models}},  other thesis, 12, 2013.

\bibitem{Coloma:2017ppo}
P.~Coloma, P.~A.~N. Machado, I.~Martinez-Soler and I.~M. Shoemaker,
  \emph{{Double-Cascade Events from New Physics in Icecube}},
  \href{http://dx.doi.org/10.1103/PhysRevLett.119.201804}{\emph{Phys. Rev.
  Lett.} {\bf 119} (2017) 201804}, [\href{http://arxiv.org/abs/1707.08573}{{\tt
  1707.08573}}].

\bibitem{Abazajian:2017tcc}
K.~N. Abazajian, \emph{{Sterile neutrinos in cosmology}},
  \href{http://dx.doi.org/10.1016/j.physrep.2017.10.003}{\emph{Phys. Rept.}
  {\bf 711-712} (2017) 1--28}, [\href{http://arxiv.org/abs/1705.01837}{{\tt
  1705.01837}}].

\bibitem{Shoemaker:2018vii}
I.~M. Shoemaker and J.~Wyenberg, \emph{{Direct Detection Experiments at the
  Neutrino Dipole Portal Frontier}},
  \href{http://dx.doi.org/10.1103/PhysRevD.99.075010}{\emph{Phys. Rev. D} {\bf
  99} (2019) 075010}, [\href{http://arxiv.org/abs/1811.12435}{{\tt
  1811.12435}}].

\bibitem{Brdar:2020quo}
V.~Brdar, A.~Greljo, J.~Kopp and T.~Opferkuch, \emph{{The Neutrino Magnetic
  Moment Portal: Cosmology, Astrophysics, and Direct Detection}},
  \href{http://dx.doi.org/10.1088/1475-7516/2021/01/039}{\emph{JCAP} {\bf 01}
  (2021) 039}, [\href{http://arxiv.org/abs/2007.15563}{{\tt 2007.15563}}].

\bibitem{Plestid:2020vqf}
R.~Plestid, \emph{{Luminous solar neutrinos I: Dipole portals}},
  \href{http://dx.doi.org/10.1103/PhysRevD.104.075027}{\emph{Phys. Rev. D} {\bf
  104} (2021) 075027}, [\href{http://arxiv.org/abs/2010.04193}{{\tt
  2010.04193}}].

\bibitem{Jodlowski:2020vhr}
K.~Jod\l{}owski and S.~Trojanowski, \emph{{Neutrino beam-dump experiment with
  FASER at the LHC}},
  \href{http://dx.doi.org/10.1007/JHEP05(2021)191}{\emph{JHEP} {\bf 05} (2021)
  191}, [\href{http://arxiv.org/abs/2011.04751}{{\tt 2011.04751}}].

\bibitem{Schwetz:2020xra}
T.~Schwetz, A.~Zhou and J.-Y. Zhu, \emph{{Constraining active-sterile neutrino
  transition magnetic moments at DUNE near and far detectors}},
  \href{http://dx.doi.org/10.1007/JHEP07(2021)200}{\emph{JHEP} {\bf 21} (2020)
  200}, [\href{http://arxiv.org/abs/2105.09699}{{\tt 2105.09699}}].

\bibitem{Ismail:2021dyp}
A.~Ismail, S.~Jana and R.~M. Abraham, \emph{{Neutrino up-scattering via the
  dipole portal at forward LHC detectors}},
  \href{http://dx.doi.org/10.1103/PhysRevD.105.055008}{\emph{Phys. Rev. D} {\bf
  105} (2022) 055008}, [\href{http://arxiv.org/abs/2109.05032}{{\tt
  2109.05032}}].

\bibitem{Miranda:2021kre}
O.~G. Miranda, D.~K. Papoulias, O.~Sanders, M.~T\'ortola and J.~W.~F. Valle,
  \emph{{Low-energy probes of sterile neutrino transition magnetic moments}},
  \href{http://dx.doi.org/10.1007/JHEP12(2021)191}{\emph{JHEP} {\bf 12} (2021)
  191}, [\href{http://arxiv.org/abs/2109.09545}{{\tt 2109.09545}}].

\bibitem{Dasgupta:2021fpn}
A.~Dasgupta, S.~K. Kang and J.~E. Kim, \emph{{Probing neutrino dipole portal at
  COHERENT experiment}},
  \href{http://dx.doi.org/10.1007/JHEP11(2021)120}{\emph{JHEP} {\bf 11} (2021)
  120}, [\href{http://arxiv.org/abs/2108.12998}{{\tt 2108.12998}}].

\bibitem{Atkinson:2021rnp}
M.~Atkinson, P.~Coloma, I.~Martinez-Soler, N.~Rocco and I.~M. Shoemaker,
  \emph{{Heavy neutrino searches through double-bang events at
  Super-Kamiokande, DUNE, and Hyper-Kamiokande}},
  \href{http://dx.doi.org/10.1007/JHEP04(2022)174}{\emph{JHEP} {\bf 04} (2022)
  174}, [\href{http://arxiv.org/abs/2105.09357}{{\tt 2105.09357}}].

\bibitem{Kamp:2022bpt}
N.~W. Kamp, M.~Hostert, A.~Schneider, S.~Vergani, C.~A. Arg\"uelles, J.~M.
  Conrad et~al., \emph{{Dipole-Coupled Neutrissimo Explanations of the
  MiniBooNE Excess Including Constraints from MINERvA Data}},
  \href{http://arxiv.org/abs/2206.07100}{{\tt 2206.07100}}.

\bibitem{Gustafson:2022rsz}
R.~A. Gustafson, R.~Plestid and I.~M. Shoemaker, \emph{{Neutrino portals,
  terrestrial upscattering, and atmospheric neutrinos}},
  \href{http://dx.doi.org/10.1103/PhysRevD.106.095037}{\emph{Phys. Rev. D} {\bf
  106} (2022) 095037}, [\href{http://arxiv.org/abs/2205.02234}{{\tt
  2205.02234}}].

\bibitem{Huang:2022pce}
G.-y. Huang, S.~Jana, M.~Lindner and W.~Rodejohann, \emph{{Probing Heavy
  Sterile Neutrinos at Ultrahigh Energy Neutrino Telescopes via the Dipole
  Portal}},  \href{http://arxiv.org/abs/2204.10347}{{\tt 2204.10347}}.

\bibitem{Li:2022bqr}
Y.-F. Li and S.-y. Xia, \emph{{Probing neutrino magnetic moments and the
  Xenon1T excess with coherent elastic solar neutrino scattering}},
  \href{http://dx.doi.org/10.1103/PhysRevD.106.095022}{\emph{Phys. Rev. D} {\bf
  106} (2022) 095022}, [\href{http://arxiv.org/abs/2203.16525}{{\tt
  2203.16525}}].

\bibitem{Acero:2022wqg}
M.~A. Acero et~al., \emph{{White Paper on Light Sterile Neutrino Searches and
  Related Phenomenology}},  \href{http://arxiv.org/abs/2203.07323}{{\tt
  2203.07323}}.

\bibitem{Feng:2022inv}
J.~L. Feng et~al., \emph{{The Forward Physics Facility at the High-Luminosity
  LHC}},  \href{http://arxiv.org/abs/2203.05090}{{\tt 2203.05090}}.

\bibitem{Hati:2022tfo}
C.~Hati, P.~Bolton, F.~F. Deppisch, K.~Fridell, J.~Harz and S.~Kulkarni,
  \emph{{Distinguishing Dirac vs Majorana Neutrinos at CE$\nu$NS experiments}},
  \href{http://dx.doi.org/10.22323/1.398.0225}{\emph{PoS} {\bf EPS-HEP2021}
  (2022) 225}.

\bibitem{Mathur:2021trm}
V.~Mathur, I.~M. Shoemaker and Z.~Tabrizi, \emph{{Using DUNE to shed light on
  the electromagnetic properties of neutrinos}},
  \href{http://dx.doi.org/10.1007/JHEP10(2022)041}{\emph{JHEP} {\bf 10} (2022)
  041}, [\href{http://arxiv.org/abs/2111.14884}{{\tt 2111.14884}}].

\bibitem{Bolton:2021pey}
P.~D. Bolton, F.~F. Deppisch, K.~Fridell, J.~Harz, C.~Hati and S.~Kulkarni,
  \emph{{Probing active-sterile neutrino transition magnetic moments with
  photon emission from CE\ensuremath{\nu}NS}},
  \href{http://dx.doi.org/10.1103/PhysRevD.106.035036}{\emph{Phys. Rev. D} {\bf
  106} (2022) 035036}, [\href{http://arxiv.org/abs/2110.02233}{{\tt
  2110.02233}}].

\bibitem{Ovchynnikov:2022rqj}
M.~Ovchynnikov, T.~Schwetz and J.-Y. Zhu, \emph{{Dipole portal and
  neutrinophilic scalars at DUNE revisited: the importance of the high-energy
  neutrino tail}},  \href{http://arxiv.org/abs/2210.13141}{{\tt 2210.13141}}.

\bibitem{Zhang:2022spf}
Y.~Zhang, M.~Song, R.~Ding and L.~Chen, \emph{{Neutrino dipole portal at
  electron colliders}},
  \href{http://dx.doi.org/10.1016/j.physletb.2022.137116}{\emph{Phys. Lett. B}
  {\bf 829} (2022) 137116}, [\href{http://arxiv.org/abs/2204.07802}{{\tt
  2204.07802}}].

\bibitem{Zhang:2023nxy}
Y.~Zhang and W.~Liu, \emph{{Probing active-sterile neutrino transition magnetic
  moments at LEP and CEPC}},  \href{http://arxiv.org/abs/2301.06050}{{\tt
  2301.06050}}.

\bibitem{Ovchynnikov:2023wgg}
M.~Ovchynnikov and J.-Y. Zhu, \emph{{Search for the dipole portal of heavy
  neutral leptons at future colliders}},
  \href{http://arxiv.org/abs/2301.08592}{{\tt 2301.08592}}.

\bibitem{Guo:2023bpo}
S.-Y. Guo, M.~Khlopov, L.~Wu and B.~Zhu, \emph{{Can Sterile Neutrino Explain
  Very High Energy Photons from GRB221009A?}},
  \href{http://arxiv.org/abs/2301.03523}{{\tt 2301.03523}}.

\bibitem{FASER:2018eoc}
{\scshape FASER} collaboration, A.~Ariga et~al., \emph{{FASER\textquoteright{}s
  physics reach for long-lived particles}},
  \href{http://dx.doi.org/10.1103/PhysRevD.99.095011}{\emph{Phys. Rev. D} {\bf
  99} (2019) 095011}, [\href{http://arxiv.org/abs/1811.12522}{{\tt
  1811.12522}}].

\bibitem{Pinfold:2019nqj}
J.~L. Pinfold, \emph{{The MoEDAL Experiment at the LHC\textemdash{}A Progress
  Report}}, \href{http://dx.doi.org/10.3390/universe5020047}{\emph{Universe}
  {\bf 5} (2019) 47}.

\bibitem{Acharya:2022nik}
B.~Acharya et~al., \emph{{MoEDAL-MAPP, an LHC Dedicated Detector Search
  Facility}},  in \emph{{2022 Snowmass Summer Study}}, 9, 2022.
\newblock \href{http://arxiv.org/abs/2209.03988}{{\tt 2209.03988}}.

\bibitem{Cerci:2021nlb}
S.~Cerci et~al., \emph{{FACET: A new long-lived particle detector in the very
  forward region of the CMS experiment}},
  \href{http://dx.doi.org/10.1007/JHEP06(2022)110}{\emph{JHEP} {\bf 2022}
  (2022) 110}, [\href{http://arxiv.org/abs/2201.00019}{{\tt 2201.00019}}].

\bibitem{Racco:2015dxa}
D.~Racco, A.~Wulzer and F.~Zwirner, \emph{{Robust collider limits on
  heavy-mediator Dark Matter}},
  \href{http://dx.doi.org/10.1007/JHEP05(2015)009}{\emph{JHEP} {\bf 05} (2015)
  009}, [\href{http://arxiv.org/abs/1502.04701}{{\tt 1502.04701}}].

\bibitem{Atre:2009rg}
A.~Atre, T.~Han, S.~Pascoli and B.~Zhang, \emph{{The Search for Heavy Majorana
  Neutrinos}},
  \href{http://dx.doi.org/10.1088/1126-6708/2009/05/030}{\emph{JHEP} {\bf 05}
  (2009) 030}, [\href{http://arxiv.org/abs/0901.3589}{{\tt 0901.3589}}].

\bibitem{Bondarenko:2018ptm}
K.~Bondarenko, A.~Boyarsky, D.~Gorbunov and O.~Ruchayskiy, \emph{{Phenomenology
  of GeV-scale Heavy Neutral Leptons}},
  \href{http://dx.doi.org/10.1007/JHEP11(2018)032}{\emph{JHEP} {\bf 11} (2018)
  032}, [\href{http://arxiv.org/abs/1805.08567}{{\tt 1805.08567}}].

\bibitem{Alloul:2013bka}
A.~Alloul, N.~D. Christensen, C.~Degrande, C.~Duhr and B.~Fuks,
  \emph{{FeynRules 2.0 - A complete toolbox for tree-level phenomenology}},
  \href{http://dx.doi.org/10.1016/j.cpc.2014.04.012}{\emph{Comput. Phys.
  Commun.} {\bf 185} (2014) 2250--2300},
  [\href{http://arxiv.org/abs/1310.1921}{{\tt 1310.1921}}].

\bibitem{Degrande:2011ua}
C.~Degrande, C.~Duhr, B.~Fuks, D.~Grellscheid, O.~Mattelaer and T.~Reiter,
  \emph{{UFO - The Universal FeynRules Output}},
  \href{http://dx.doi.org/10.1016/j.cpc.2012.01.022}{\emph{Comput. Phys.
  Commun.} {\bf 183} (2012) 1201--1214},
  [\href{http://arxiv.org/abs/1108.2040}{{\tt 1108.2040}}].

\bibitem{Alwall:2014hca}
J.~Alwall, R.~Frederix, S.~Frixione, V.~Hirschi, F.~Maltoni, O.~Mattelaer
  et~al., \emph{{The automated computation of tree-level and next-to-leading
  order differential cross sections, and their matching to parton shower
  simulations}}, \href{http://dx.doi.org/10.1007/JHEP07(2014)079}{\emph{JHEP}
  {\bf 07} (2014) 079}, [\href{http://arxiv.org/abs/1405.0301}{{\tt
  1405.0301}}].

\bibitem{Sjostrand:2014zea}
T.~Sj\"ostrand, S.~Ask, J.~R. Christiansen, R.~Corke, N.~Desai, P.~Ilten
  et~al., \emph{{An introduction to PYTHIA 8.2}},
  \href{http://dx.doi.org/10.1016/j.cpc.2015.01.024}{\emph{Comput. Phys.
  Commun.} {\bf 191} (2015) 159--177},
  [\href{http://arxiv.org/abs/1410.3012}{{\tt 1410.3012}}].

\bibitem{deFavereau:2013fsa}
{\scshape DELPHES 3} collaboration, J.~de~Favereau, C.~Delaere, P.~Demin,
  A.~Giammanco, V.~Lema\^\i{}tre, A.~Mertens et~al., \emph{{DELPHES 3, A
  modular framework for fast simulation of a generic collider experiment}},
  \href{http://dx.doi.org/10.1007/JHEP02(2014)057}{\emph{JHEP} {\bf 02} (2014)
  057}, [\href{http://arxiv.org/abs/1307.6346}{{\tt 1307.6346}}].

\bibitem{Cacciari:2011ma}
M.~Cacciari, G.~P. Salam and G.~Soyez, \emph{{FastJet User Manual}},
  \href{http://dx.doi.org/10.1140/epjc/s10052-012-1896-2}{\emph{Eur. Phys. J.
  C} {\bf 72} (2012) 1896}, [\href{http://arxiv.org/abs/1111.6097}{{\tt
  1111.6097}}].

\bibitem{Alimena:2019zri}
J.~Alimena et~al., \emph{{Searching for long-lived particles beyond the
  Standard Model at the Large Hadron Collider}},
  \href{http://dx.doi.org/10.1088/1361-6471/ab4574}{\emph{J. Phys. G} {\bf 47}
  (2020) 090501}, [\href{http://arxiv.org/abs/1903.04497}{{\tt 1903.04497}}].

\bibitem{Gligorov:2018vkc}
V.~V. Gligorov, S.~Knapen, B.~Nachman, M.~Papucci and D.~J. Robinson,
  \emph{{Leveraging the ALICE/L3 cavern for long-lived particle searches}},
  \href{http://dx.doi.org/10.1103/PhysRevD.99.015023}{\emph{Phys. Rev. D} {\bf
  99} (2019) 015023}, [\href{http://arxiv.org/abs/1810.03636}{{\tt
  1810.03636}}].

\bibitem{Bauer:2019vqk}
M.~Bauer, O.~Brandt, L.~Lee and C.~Ohm, \emph{{ANUBIS: Proposal to search for
  long-lived neutral particles in CERN service shafts}},
  \href{http://arxiv.org/abs/1909.13022}{{\tt 1909.13022}}.

\bibitem{Gligorov:2017nwh}
V.~V. Gligorov, S.~Knapen, M.~Papucci and D.~J. Robinson, \emph{{Searching for
  Long-lived Particles: A Compact Detector for Exotics at LHCb}},
  \href{http://dx.doi.org/10.1103/PhysRevD.97.015023}{\emph{Phys. Rev. D} {\bf
  97} (2018) 015023}, [\href{http://arxiv.org/abs/1708.09395}{{\tt
  1708.09395}}].

\bibitem{Curtin:2018mvb}
D.~Curtin et~al., \emph{{Long-Lived Particles at the Energy Frontier: The
  MATHUSLA Physics Case}},
  \href{http://dx.doi.org/10.1088/1361-6633/ab28d6}{\emph{Rept. Prog. Phys.}
  {\bf 82} (2019) 116201}, [\href{http://arxiv.org/abs/1806.07396}{{\tt
  1806.07396}}].

\bibitem{FASER:2018bac}
{\scshape FASER} collaboration, A.~Ariga et~al., \emph{{Technical Proposal for
  FASER: ForwArd Search ExpeRiment at the LHC}},
  \href{http://arxiv.org/abs/1812.09139}{{\tt 1812.09139}}.

\bibitem{CHARM-II:1989srx}
{\scshape CHARM-II} collaboration, D.~Geiregat et~al., \emph{{A New
  Determination of the Electroweak Mixing Angle From $\nu_\mu$ Electron
  Scattering}},
  \href{http://dx.doi.org/10.1016/0370-2693(89)90457-7}{\emph{Phys. Lett. B}
  {\bf 232} (1989) 539}.

\bibitem{LSND:1996ubh}
{\scshape LSND} collaboration, C.~Athanassopoulos et~al., \emph{{Evidence for
  anti-muon-neutrino ---\ensuremath{>} anti-electron-neutrino oscillations from
  the LSND experiment at LAMPF}},
  \href{http://dx.doi.org/10.1103/PhysRevLett.77.3082}{\emph{Phys. Rev. Lett.}
  {\bf 77} (1996) 3082--3085},
  [\href{http://arxiv.org/abs/nucl-ex/9605003}{{\tt nucl-ex/9605003}}].

\bibitem{MiniBooNE:2007uho}
{\scshape MiniBooNE} collaboration, A.~A. Aguilar-Arevalo et~al., \emph{{A
  Search for Electron Neutrino Appearance at the $\Delta m^2 \sim 1 eV^2$
  Scale}}, \href{http://dx.doi.org/10.1103/PhysRevLett.98.231801}{\emph{Phys.
  Rev. Lett.} {\bf 98} (2007) 231801},
  [\href{http://arxiv.org/abs/0704.1500}{{\tt 0704.1500}}].

\bibitem{Vannucci:2014wna}
F.~Vannucci, \emph{{The NOMAD Experiment at CERN}},
  \href{http://dx.doi.org/10.1155/2014/129694}{\emph{Adv. High Energy Phys.}
  {\bf 2014} (2014) 129694}.

\bibitem{NOMAD:1997pcg}
{\scshape NOMAD} collaboration, J.~Altegoer et~al., \emph{{The NOMAD experiment
  at the CERN SPS}},
  \href{http://dx.doi.org/10.1016/S0168-9002(97)01079-6}{\emph{Nucl. Instrum.
  Meth. A} {\bf 404} (1998) 96--128}.

\bibitem{NOMAD:1998pxi}
{\scshape NOMAD} collaboration, J.~Altegoer et~al., \emph{{Search for a new
  gauge boson in pi0 decays}},
  \href{http://dx.doi.org/10.1016/S0370-2693(98)00402-X}{\emph{Phys. Lett. B}
  {\bf 428} (1998) 197--205}, [\href{http://arxiv.org/abs/hep-ex/9804003}{{\tt
  hep-ex/9804003}}].

\bibitem{OPAL:1994kgw}
{\scshape OPAL} collaboration, R.~Akers et~al., \emph{{Measurement of single
  photon production in e+ e- collisions near the Z0 resonance}},
  \href{http://dx.doi.org/10.1007/BF01571303}{\emph{Z. Phys. C} {\bf 65} (1995)
  47--66}.

\bibitem{L3:1997exg}
{\scshape L3} collaboration, M.~Acciarri et~al., \emph{{Search for new physics
  in energetic single photon production in $e^{+} e^{-}$ annihilation at the
  $Z$ resonance}},
  \href{http://dx.doi.org/10.1016/S0370-2693(97)01003-4}{\emph{Phys. Lett. B}
  {\bf 412} (1997) 201--209}.

\bibitem{ATLAS:2017nga}
{\scshape ATLAS} collaboration, M.~Aaboud et~al., \emph{{Search for dark matter
  at $\sqrt{s}=13$ TeV in final states containing an energetic photon and large
  missing transverse momentum with the ATLAS detector}},
  \href{http://dx.doi.org/10.1140/epjc/s10052-017-4965-8}{\emph{Eur. Phys. J.
  C} {\bf 77} (2017) 393}, [\href{http://arxiv.org/abs/1704.03848}{{\tt
  1704.03848}}].

\bibitem{CMS:2015loa}
{\scshape CMS} collaboration, V.~Khachatryan et~al., \emph{{Search for
  supersymmetry in events with a photon, a lepton, and missing transverse
  momentum in pp collisions at $\sqrt s=$ 8 TeV}},
  \href{http://dx.doi.org/10.1016/j.physletb.2016.03.039}{\emph{Phys. Lett. B}
  {\bf 757} (2016) 6--31}, [\href{http://arxiv.org/abs/1508.01218}{{\tt
  1508.01218}}].

\bibitem{ATLAS:2020uiq}
{\scshape ATLAS} collaboration, G.~Aad et~al., \emph{{Search for dark matter in
  association with an energetic photon in $pp$ collisions at $\sqrt{s}$ = 13
  TeV with the ATLAS detector}},
  \href{http://dx.doi.org/10.1007/JHEP02(2021)226}{\emph{JHEP} {\bf 02} (2021)
  226}, [\href{http://arxiv.org/abs/2011.05259}{{\tt 2011.05259}}].

\bibitem{CMS:2018fon}
{\scshape CMS} collaboration, A.~M. Sirunyan et~al., \emph{{Search for
  supersymmetry in events with a photon, a lepton, and missing transverse
  momentum in proton-proton collisions at $\sqrt{s} =$ 13 TeV}},
  \href{http://dx.doi.org/10.1007/JHEP01(2019)154}{\emph{JHEP} {\bf 01} (2019)
  154}, [\href{http://arxiv.org/abs/1812.04066}{{\tt 1812.04066}}].

\bibitem{Kamiokande-II:1987idp}
{\scshape Kamiokande-II} collaboration, K.~Hirata et~al., \emph{{Observation of
  a Neutrino Burst from the Supernova SN 1987a}},
  \href{http://dx.doi.org/10.1103/PhysRevLett.58.1490}{\emph{Phys. Rev. Lett.}
  {\bf 58} (1987) 1490--1493}.

\bibitem{Alekseev:1988gp}
E.~N. Alekseev, L.~N. Alekseeva, I.~V. Krivosheina and V.~I. Volchenko,
  \emph{{Detection of the Neutrino Signal From {SN1987A} in the {LMC} Using the
  Inr Baksan Underground Scintillation Telescope}},
  \href{http://dx.doi.org/10.1016/0370-2693(88)91651-6}{\emph{Phys. Lett. B}
  {\bf 205} (1988) 209--214}.

\bibitem{Bionta:1987qt}
R.~M. Bionta et~al., \emph{{Observation of a Neutrino Burst in Coincidence with
  Supernova SN 1987a in the Large Magellanic Cloud}},
  \href{http://dx.doi.org/10.1103/PhysRevLett.58.1494}{\emph{Phys. Rev. Lett.}
  {\bf 58} (1987) 1494}.

\end{thebibliography}\endgroup
\end{document}